\documentclass[nofootinbib,floatfix,amsmath,amssymb,showpacs,superscriptaddress,notitlepage,twocolumn]{revtex4-1}

\usepackage{rotating}
\usepackage{graphicx}
\usepackage{dcolumn}
\usepackage{bm}
\usepackage{color}
\usepackage{multirow}
\usepackage{tabularx}
\usepackage{comment}
\usepackage{wasysym}

\newcolumntype{C}{>{\centering\arraybackslash}X}
\newcolumntype{R}{>{\raggedleft\arraybackslash}X}

\begin{document}

\title{A look inside charmed-strange baryons from lattice QCD}
\author{K. U. Can}
\affiliation{Department of Physics, H-27, Tokyo Institute of Technology, Meguro, Tokyo 152-8551 Japan}
\author{G. Erkol}
\affiliation{Department of Natural and Mathematical Sciences, Faculty of Engineering, Ozyegin University, Nisantepe Mah. Orman Sok. No:34-36, Alemdag 34794 Cekmekoy, Istanbul Turkey}
\author{M. Oka}%
\affiliation{Department of Physics, H-27, Tokyo Institute of Technology, Meguro, Tokyo 152-8551 Japan}
\affiliation{Advanced Science Research Center, Japan Atomic Energy Agency, Tokai, Ibaraki, 319-1195 Japan}
\author{T. T. Takahashi}%
\affiliation{Gunma National College of Technology, Maebashi, Gunma 371-8530 Japan}

\date{\today}

\begin{abstract}
The electromagnetic form factors of the spin-3/2 $\Omega$ baryons, namely $\Omega$, $\Omega_c^\ast$, $\Omega_{cc}^\ast$ and $\Omega_{ccc}$, are calculated in full QCD on $32^3\times 64$ PACS-CS lattices with a pion mass of 156(9) MeV. The electric charge radii and magnetic moments from the $E0$ and $M1$ multipole form factors are extracted. Results for the electric quadrupole form factors, $E2$, are also given. Quark sector contributions are computed individually for each observable and then combined to obtain the baryon properties. We find that the charm quark contributions are systematically smaller than the strange-quark contributions in the case of the charge radii and magnetic moments. $E2$ moments of the $\Omega_{cc}^\ast$ and $\Omega_{ccc}$ provide a statistically significant data to conclude that their electric charge distributions are deformed to an oblate shape. Properties of the spin-1/2 $\Omega_c$ and $\Omega_{cc}$ baryons are also computed and a thorough comparison is given. This complete study gives valuable hints about the heavy-quark dynamics in charmed hadrons.
\end{abstract}
\maketitle

\section{Introduction}
     
Since the discovery of the proton's internal structure, form factors have been the tools to investigate the inner structure of hadrons. One of the intriguing properties of hadrons is their electromagnetic structure, such as their charge radii and magnetic moments. There have been enormous efforts to determine the electromagnetic form factors both experimentally and theoretically. On the theoretical side, lattice QCD is a widely used first-principles calculation framework to study these form factors. Main challenges for the lattice QCD form factor calculations have been the pseudoscalar/vector-meson states and the nucleon (see \cite{Constantinou:2014tga} for a review) with only recent works probing the octet~\cite{PhysRevD.74.093005, Shanahan:2014uka, Shanahan:2014cga} and decuplet sectors~\cite{PhysRevD.80.054505, PhysRevD.80.054505, Alexandrou:2008bn, Alexandrou:2010jv}. 

Improvement in our understanding of the light hadron electromagnetic structure makes it timely to study the heavy-flavor hadrons further. Comparison of the two sectors would reveal differences in the quark-gluon dynamics of heavy flavors. We have recently examined the internal structure and the quark dynamics of the hadrons that contain at least one charm quark using lattice QCD. Initially in the meson sector, we found that the electric charge radii and the magnetic moments of the $D$ and $D^\ast$ mesons to be smaller~\cite{Can:2012tx} as compared to those of the $\pi$ and $\rho$ mesons. An investigation of the quark contributions revealed that the decrease is mainly due to the smallness of the charm-quark contributions to the observables. In this same vein, we extended our calculations to the spin-1/2 light-charmed and strange-charmed baryons. We observed a similar behavior in the internal structure caused by the heavy charm quark~\cite{Can:2013zpa,Can:2013tna}. Finally, we made a lattice study of the experimentally observed $\Omega_c^\ast \left(\frac{3}{2}^+\right) \rightarrow \Omega_c \left(\frac{1}{2}^+\right) \gamma$ radiative transition and estimated its lifetime~\cite{Bahtiyar2015281}.    

In this work, our aim is to broaden our perspective by including the elastic electromagnetic form factors of the $J=\frac{3}{2}^+$ strange-charmed baryons to examine the spin-dependence of the quark dynamics. We compute the electromagnetic form factors of the $\Omega$, $\Omega_c^\ast$, $\Omega_{cc}^\ast$ and $\Omega_{ccc}$ baryons as well as the $\Omega_c$, $\Omega_{cc}$ baryons with $J=\frac{1}{2}^+$. We extract the electric charge radii and the magnetic moments, and give a thorough comparison of both for the spin-3/2 and spin-1/2 sectors, which helps to improve our understanding of the nonperturbative structure of the strange-charmed baryons.    

Our work is organised as follows: In Sec.~\ref{lat_for} we outline the continuum and the lattice formulation to calculate the electromagnetic form factors of the spin-3/2 baryons and discuss the lattice setup and the methods we utilise. Sec.~\ref{res_dis} is devoted to the results. We present our results for the baryon masses, electric charge radii and magnetic moments of the spin-3/2 baryons as well as the results for their spin-1/2 counterparts. The electric quadrupole moments of the spin-3/2 baryons are also given in this section. Our main findings are itemised at the end of each subsection and a summary is given in Sec.~\ref{conc}.   

\section{Theoretical formulation and lattice setup}
\label{lat_for}
Electromagnetic form factors of baryons can be calculated through their matrix elements of the electromagnetic vector current $j_\mu=\sum_q e_q \bar{q}(x)\gamma_\mu q(x)$, where $q$ runs over the quark content of the baryon in consideration. We refer the reader to Ref.~\cite{Can:2013tna} for the analytical expressions of spin-1/2 baryons. Here we only give the details for spin-3/2 baryons.
\subsection{Theoretical formulation}
The electromagnetic transition matrix element for the spin-3/2 baryons can be written as 
\begin{align}\label{matrix}
	\begin{split}
	\langle \mathcal{B}_\sigma(p',s') | j_\mu| &\mathcal{B}_\tau(p,s) \rangle 
	\\&= \sqrt{\frac{M_\mathcal{B}^2}{E\,E^\prime}} \bar{u}_\sigma(p',s') \mathcal{O}^{\sigma\mu\tau} u_\tau(p,s),
	\end{split}
\end{align}
where $p(s)$ and $p'(s')$ denote the four momentum (spin) of the initial and final states, respectively. $M_\mathcal{B}$ is the mass of the baryon, $E$ ($E^\prime$) is the energy of the incoming (outgoing) baryon state and $u_\alpha(p,s)$ is the baryon spinor in the Rarita-Schwinger formalism. The tensor in Eq.~\eqref{matrix} can be written in a Lorentz-covariant form as~\cite{Nozawa:1990gt}
\begin{align}
	\begin{split}
		\mathcal{O}^{\sigma\mu\tau} = &-g^{\sigma\tau} \left\{ a_1 \gamma^\mu + \frac{a_2}{2 M_\mathcal{B}} P^\mu \right\} \\
																	&- \frac{q^\sigma q^\tau}{(2 M_\mathcal{B})^2} \left\{ c_1 \gamma^\mu + \frac{c_2}{2 M_\mathcal{B}} P^\mu \right\},
	\end{split}
\end{align}  
where $P = p + p'$ and $q = p' - p$. The multipole form factors are defined in terms of the covariant vertex functions $a_1,\,a_2,\,c_1$ and $c_2$ as,
\begin{align}
	\begin{split}		
		G_{E0}(q^2) &= (1+\frac{2}{3}\tau) \left\{ a_1 + (1 + \tau) a_2 \right\} \\
								&\quad- \frac{1}{3}\tau (1+\tau) \left\{c_1 + (1+\tau) c_2 \right\},
	\end{split}
\\
	\begin{split}
			G_{E2}(q^2) &=  \left\{ a_1 + (1 + \tau) a_2 \right\} \\
								&\quad- \frac{1}{2}(1+\tau) \left\{c_1 + (1+\tau) c_2 \right\},
	\end{split}
\\
	G_{M1}(q^2) &= (1 + \frac{4}{3}\tau)a_1 - \frac{2}{3}\tau(1+\tau)c_1, 
\\ 
	G_{M3}(q^2)& = a_1 - \frac{1}{2}(1+\tau)c_1,
\end{align}
with $\tau = -q^2/(2 M_B)^2$. These multipole form factors are referred to as electric-charge ($E0$), electric-quadrupole ($E2$), magnetic-dipole ($M1$) and magnetic-octupole (M3) multipole form factors.

The two- and three-point correlation functions for spin-3/2 baryons are defined as,
\allowdisplaybreaks{
\begin{align}
	\begin{split}\label{deltacf}
	\langle G_{\sigma\tau}^{\mathcal{BB}}(t; {\bf p};\Gamma_4)\rangle &=\sum_{\bf x}e^{-i{\bf p}\cdot {\bf x}}\Gamma_4^{\alpha\alpha^\prime} \\
	&\times \langle \text{vac} | T [\eta_{\sigma}^\alpha(x) \bar{\eta}_{\tau}^{\alpha^\prime}(0)] | \text{vac}\rangle,
	\end{split}
\end{align}	
\begin{align}	
	\begin{split}\label{thrpcf}
	\langle G_{\sigma\tau}^{\mathcal{B} j^\mu \mathcal{B}}(t_2,t_1; {\bf p}^\prime, {\bf p};\mathbf{\Gamma})\rangle =-i\sum_{{\bf x_2},{\bf x_1}} e^{-i{\bf p}^\prime \cdot {\bf x_2}} e^{i{\bf q}\cdot {\bf x_1}}  & \\
	\times \mathbf{\Gamma}^{\alpha\alpha^\prime} \langle \text{vac} | T [\eta_\sigma^\alpha(x_2) j_\mu(x_1) \bar{\eta}_\tau^{\alpha^\prime}(0)] | \text{vac}\rangle,&
	\end{split}
\end{align}
}%
with the spin projection matrices 
\begin{equation}
	\Gamma_i=\frac{1}{2}\left(\begin{matrix}\sigma_i & 0 \\ 0 & 0 \end{matrix}\right), \qquad \Gamma_4=\frac{1}{2}\left(\begin{matrix}I & 0 \\ 0 & 0 \end{matrix}\right),
\end{equation}
where $\sigma_i$ are the Pauli spin matrices, $\alpha$, $\beta$ denote the Dirac indices and $\sigma$, $\tau$ are the Lorentz indices of the spin-3/2 interpolating fields. The baryon interpolating fields are chosen, similarly to those of Delta baryon as
\begin{align}
	\begin{split}
		\eta_\mu(x)=\frac{1}{\sqrt{3}}\epsilon^{ijk} \{2[q_1^{T i}(x) C \gamma_\mu q_2^j(x)]q_3^k(x) & \\
		+[q_1^{T i}(x) C \gamma_\mu q_3^j(x)]q_2^k(x)\}&,
		\label{deltaint}
	\end{split}
\end{align}
where $i$, $j$, $k$ denote the color indices and $C=\gamma_4\gamma_2$. $q_1$, $q_2$, $q_3$ are the quark flavors and chosen as $(q_1,q_2,q_3)=$$\{(s,s,s)$, $(s,s,c)$, $(s,c,c)$, $(c,c,c)\}$ for $\Omega$, $\Omega_c^\ast$, $\Omega_{cc}^\ast$ and $\Omega_{ccc}$ baryons, respectively. It has been shown in Refs.~\cite{Alexandrou:2014sha,PhysRevD.80.054505} that the interpolating field in Eq.~\eqref{deltaint} has minimal overlap with spin-1/2 states and therefore spin-3/2 projection is not necessary. 

Inserting a complete set of eigenstates $\sum_s |(p,s)\rangle\langle(p,s)|$ into \eqref{deltacf} and \eqref{thrpcf} and taking the large Euclidean time limit, $t_2-t_1$ and $t_1\gg a$, correlation functions reduce to
\begin{align}
	\begin{split}\label{deltacf_el}	
		&\langle G_{\sigma\tau}^{\mathcal{BB}}(t; {\bf p};\Gamma_4)\rangle = \\
		&\qquad Z_\mathcal{B}(p) \bar{Z}_\mathcal{B}(p) \frac{M_\mathcal{B}}{E} e^{- E t} \rm{Tr}[\Gamma_4 \Lambda_{\sigma\tau}],
	\end{split}
\\
	\begin{split}\label{thrpcf_el}
		&\langle G_{\sigma\tau}^{\mathcal{B} j^\mu \mathcal{B}}(t_2,t_1; {\bf p}^\prime, {\bf p};\mathbf{\Gamma})\rangle = Z_\mathcal{B}(p^\prime) \bar{Z}_\mathcal{B}(p) \frac{M_\mathcal{B}^2}{E E^\prime}  \\
		&\qquad\times e^{-E^\prime (t_2 - t_1)} e^{-E t_1} \rm{Tr}[\mathbf{\Gamma} \Lambda_{\sigma \sigma^\prime}(p^\prime) \mathcal{O}^{\sigma^\prime \mu \tau^\prime} \Lambda_{\tau^\prime \tau}(p)] ,
	\end{split}
\end{align}
where the trace acts in the Dirac space, the $Z_\mathcal{B}(p)$ is the overlap factor of the interpolating field to the corresponding baryon state and $\Lambda_{\sigma \tau}$ is the Rarita-Schwinger spin sum for the spin-3/2 field in Euclidean space, defined as 
\begin{align}
	\begin{split}
		&\sum_s u_\sigma(p,s) \bar{u}_\tau(p,s) =\frac{-i\gamma\cdot p+M_\mathcal{B}}{2M_\mathcal{B}}\\
		&\times\left[g_{\sigma\tau}-\frac{1}{3}\gamma_\sigma \gamma_\tau +\frac{2p_\sigma p_\tau}{3M_\mathcal{B}^2}-i\frac{p_\sigma \gamma_\tau-p_\tau \gamma_\sigma}{3M_\mathcal{B}}\right] \\
		& \equiv \Lambda_{\sigma \tau}(p).
	\end{split}
\end{align} 

To extract the multipole form factors we consider the following ratio of the correlation functions given in Eqs.~\eqref{deltacf} and \eqref{thrpcf},
\begin{align}
	\label{plato_ratio}
	\begin{split}
		& R_{\sigma\;\;\,\tau}^{\;\;\,\mu}(t_2,t_1;\mathbf{p'},\mathbf{p};\mathbf{\Gamma}) = \\
		& \left[\frac{\langle G_{\sigma\tau}^{\mathcal{B} j^\mu \mathcal{B}}(t_2,t_1; {\bf p}^\prime, {\bf p};\mathbf{\Gamma})\rangle \langle G_{\sigma\tau}^{\mathcal{B} j^\mu \mathcal{B}}(t_2,t_1; {\bf p}, -{\bf p}^\prime;\mathbf{\Gamma})\rangle}{\langle G_{\sigma\tau}^{\mathcal{BB}}(t_2; {\bf p}^\prime;\Gamma_4)\rangle \langle G_{\sigma\tau}^{\mathcal{BB}}(t_2; -{\bf p};\Gamma_4)\rangle}\right]^{1/2} \\
		& \xrightarrow[t_2-t_1\gg a]{t_1\gg a} \left(\frac{E_p + M_\mathcal{B}}{2E_p} \right)^{1/2} \left(\frac{E_{p^\prime} + M_\mathcal{B}}{2E_{p^\prime}} \right)^{1/2} \\
		& \qquad\times \Pi_{\sigma\;\;\,\tau}^{\;\;\,\mu}({\bf p^\prime},{\bf p};\mathbf{\Gamma}).
	\end{split} 
\end{align}
Note that there is no sum over the repeated indices.

The multipole form factors can be extracted by using the following combinations of $\Pi_{\sigma\;\;\,\tau}^{\;\;\,\mu}({\bf p^\prime},{\bf p};\mathbf{\Gamma})$~\cite{PhysRevD.80.054505}:
\begin{align}
	\label{E0lat}
	\begin{split}
			G_{E0}(q^2) &= \frac{1}{3} \left( \Pi_{1\;\,1}^{\;\,4}({\bf q}_i,0;\Gamma_4) +\Pi_{2\;\,2}^{\;\,4}({\bf q}_i,0;\Gamma_4) \right. \\
			& \left. + \Pi_{3\;\,3}^{\;\,4}({\bf q}_i,0;\Gamma_4)\right.),
	\end{split}
\end{align}	
\begin{align}
	\label{E2lat}
	\begin{split}
			G_{E2}(q^2) &= 2 \frac{M(E+M)}{|\mathbf{q}_i|^2} \left( \Pi_{1\;\,1}^{\;\,4}({\bf q}_i,0;\Gamma_4) \right.\\
			& \left. +\Pi_{2\;\,2}^{\;\,4}({\bf q}_i,0;\Gamma_4) - 2\Pi_{3\;\,3}^{\;\,4}({\bf q}_i,0;\Gamma_4)\right.),
	\end{split}
\end{align}	
\begin{align}
	\label{M1lat}
	\begin{split}
			G_{M1}(q^2) &= -\frac{3}{5}\frac{E+M}{|\mathbf{q}_1|^2} \left( \Pi_{1\;\,1}^{\;\,3}({\bf q}_1,0;\Gamma_2) \right.\\
			& \left. +\Pi_{2\;\,2}^{\;\,3}({\bf q}_1,0;\Gamma_2) + \Pi_{3\;\,3}^{\;\,3}({\bf q}_1,0;\Gamma_2)\right.),
	\end{split}
\end{align}	
\begin{align}
	\label{M3lat}
	\begin{split}
			G_{M3}(q^2) &= -4\frac{M(E+M)^2}{|\mathbf{q}_1|^3} \left( \Pi_{1\;\,1}^{\;\,3}({\bf q}_1,0;\Gamma_2) \right. \\
			& \left. +\Pi_{2\;\,2}^{\;\,3}({\bf q}_1,0;\Gamma_2) - \frac{3}{2} \Pi_{3\;\,3}^{\;\,3}({\bf q}_1,0;\Gamma_2) \right.),
	\end{split}
\end{align}
where $i=1,2,3$ and $\mathbf{q}_i$ are the momentum vectors in three spatial directions. In case of the $E2$ form factor, it is possible to exploit the symmetry,
\begin{equation}
	\Pi_{2\;\,2}^{\;\,4}({\bf q}_i,0;\Gamma_4) = \Pi_{3\;\,3}^{\;\,4}({\bf q}_i,0;\Gamma_4),
\end{equation}
and define an average
\begin{align}
	\begin{split}
			\Pi_{avg}^4&({\bf q}_i,0;\Gamma_4) = \\
			&\frac{1}{2} \left[ \Pi_{2\;\,2}^{\;\,4}({\bf q}_i,0;\Gamma_4) + \Pi_{3\;\,3}^{\;\,4}({\bf q}_i,0;\Gamma_4) \right],
	\end{split}
\end{align}
in order to decrease the statistical noise in $G_{E2}$. With the above definitions, $G_{E2}$ form factor can be rewritten as
\begin{align}
	\begin{split}
		G_{E2}(q^2) &= 2 \frac{M(E+M)}{|\mathbf{q}_i|^2} \left( \Pi_{1\;\,1}^{\;\,4}({\bf q}_i,0;\Gamma_4) \right. \\
		&\left. - \Pi_{avg}^4({\bf q}_i,0;\Gamma_4) \right).
	\end{split}
\end{align}

We consider an average over momentum directions for both $E0$ and $E2$ form factors. In case of the $M1$ form factor, we make a redefinition to utilise all possible index combinations in order to improve the signal. Sum of all correlation-function ratios for $M1$ is written as   
\begin{equation}
	\begin{split}
			G_{M1}(q^2) &= -\frac{3}{5}\frac{(E+M)}{|\mathbf{q}|^2} \frac{1}{6} \sum_{\substack{i,j,k=1 \\ i\neq j\neq k}}^3 \left[ \Pi_{i\;\,i}^{\;\,j}({\bf q}_i,0;\Gamma_k) \right.\\
			& \left. + \Pi_{i\;\,i}^{\;\,j}({\bf q}_k,0;\Gamma_i) + \Pi_{i\;\,i}^{\;\,i}({\bf q}_j,0;\Gamma_k)\right].
	\end{split}
\end{equation}

Compared to the dominant form factors $E0$ and $M1$ we have observed that the data for the $E2$ and $M3$ form factor is much noisier. It turns out that with the limited number of gauge configurations we have at the smallest quark mass, the data for $M3$ moments are too noisy to allow a statistically significant value. Thus, we omit the $M3$ form factor in this work and extract only the $E0$, $M1$ and $E2$ form factors for the lowest allowed lattice momentum transfer.

It is possible that the higher order form factors in the expansion interfere with the leading and sub-leading form factors that we consider. Although a dedicated study would be needed to have a strong conclusion we note that the agreement between the results obtained from different lattice formulations of Ref.\cite{PhysRevD.80.054505} and Ref.\cite{Alexandrou:2010jv} suggests that the interference effects are minimal.   
\subsection{Lattice setup}

We have run our simulations on gauge configurations generated by PACS-CS collaboration~\cite{Aoki:2008sm} with the nonperturbatively $O(a)$-improved Wilson quark action and the Iwasaki gauge action. The details of the gauge configurations are given in Table~\ref{lat_det}. The simulations are carried out with near physical $u$,$d$ sea quarks of hopping parameter $\kappa^{u,d}=$ 0.13781. This corresponds to a pion mass of approximately 156~MeV~\cite{Aoki:2008sm}. The hopping parameter for the sea $s$ quark is fixed to $\kappa_\text{sea}^{s}=0.13640$ and the hopping parameter for the valence $s$-quark is taken to be the same.

\begin{table}[h]
	\caption{ The details of the gauge configurations used in this work~\cite{Aoki:2008sm}. We list the number of flavors~($N_f$), the lattice spacing~($a$), the lattice size~($L$), inverse gauge coupling~($\beta$), clover coefficient~($c_{SW}$), number of gauge configurations employed ($N_{gc}$) and the corresponding pion mass~($m_\pi$).
}
\begin{center}
	\small
	{
	\setlength{\extrarowheight}{7pt}
\begin{tabularx}{0.5\textwidth}{cccccccc}
			\hline\hline 
			$N_s^3\times N_t$  & $N_f$ & $a$ [fm] &  $L$ [fm] & $\beta$ & $c_{SW}$ & $N_{gc}$ & $m_\pi$ [MeV] \\
			\hline
			$32^3 \times 64$ & 2+1 & 0.0907(13)  & 2.90 & 1.90 & 1.715 & 194 & 156(7)(2) \\
			\hline \hline			
\end{tabularx}
	\label{lat_det}
	}
\end{center}
\end{table}

For the charm quarks, we employ the Clover action and use the hopping parameter value, $\kappa_c=0.1246$, which we have determined in our previous work~\cite{Can:2013tna}. In order to tune the hopping parameter we apply the Fermilab method~\cite{ElKhadra:1996mp} in the form employed by the Fermilab Lattice and MILC Collaborations~\cite{Burch:2009az, Bernard:2010fr}. A similar procedure has been recently used to study charmonium, heavy-light meson resonances and their scattering with pion and kaon~\cite{Mohler:2011ke, Mohler:2012na, Mohler:2013rwa}. In the Fermilab method's simplest application one sets the Clover coefficients $c_E$ and $c_B$ to the tadpole-improved value $1/u_0^3$, where $u_0$ is the average link. We follow the approach used in Ref.~\cite{Mohler:2011ke} to estimate the $u_0$ as the fourth root of the average plaquette and determine the charm-quark hopping parameter $\kappa_c$ nonperturbatively by tuning the spin-averaged static masses of charmonium and heavy-light mesons to their experimental results.

We make our simulations with the lowest allowed lattice momentum transfer $q=2\pi/(N_s a)$, where $N_s$ is the spatial dimension of the lattice and $a$ is the lattice spacing. This corresponds to three-momentum squared value of ${\bf q}^2$ $=$ $0.183$ GeV$^2$. In order to increase statistics, we insert all possible momentum components, namely $(|q_x|,|q_y|,|q_z|)$ $=$ $(-1,0,0)$, $(0,-1,0)$, $(0,0,-1)$, $(1,0,0)$, $(0,1,0)$, $(0,0,1)$. We also consider vector-current and spin projections along all spatial directions and take into account all Lorentz components of the Rarita-Schwinger field. We employ a wall-source/sink method~\cite{Can:2012tx}, which enables us to simultaneously extract all the components of the correlators given in Eqs.~(\ref{E0lat}-\ref{M1lat}), eliminating the need for extra inversions. However, since wall smearing is not a gauge-invariant smearing method, gauge fixing becomes a necessity. We choose to work with Coulomb gauged configurations, which lead to a better ground-state saturation. 

In order to improve the ground-state coupling, non-wall smeared source and sink are smeared in a gauge-invariant manner using a Gaussian form. In the case of $s$ quark, we choose the smearing parameters so as to give a root-mean-square radius of $\langle r_l \rangle \sim 0.5$~fm. We have measured the size of the charm quark charge radius to be small compared to the light and strange quarks, both in mesons~\cite{Can:2012tx} and baryons~\cite{Can:2013zpa,Can:2013tna}. Therefore, we adjust the smearing parameters to obtain $\langle r_c \rangle=\langle r_l \rangle/3$.

The source-sink time separation is fixed to 1.09 fm ($t_2=12a$), which has been shown to be sufficient to avoid excited state contaminations for electromagnetic form factors~\cite{Can:2013tna}. Using translational symmetry, we have employed multiple source-sink pairs by shifting them 12 lattice units in the temporal direction. All statistical errors are estimated by a single-elimination jackknife analysis. We calculate the connected diagrams only and consider the point-split lattice current
\begin{equation}
j_\mu = 1/2[\bar{q}(x+\mu)U^\dagger_\mu(1+\gamma_\mu)q(x) -\bar{q}(x)U_\mu(1-\gamma_\mu)q(x+\mu)],
\end{equation}
which is conserved by Wilson fermions.

\section{Results and Discussion}
\label{res_dis}
\subsection{Baryon masses}
\begin{figure}[ht]
	\centering
	\includegraphics[width=0.475\textwidth]{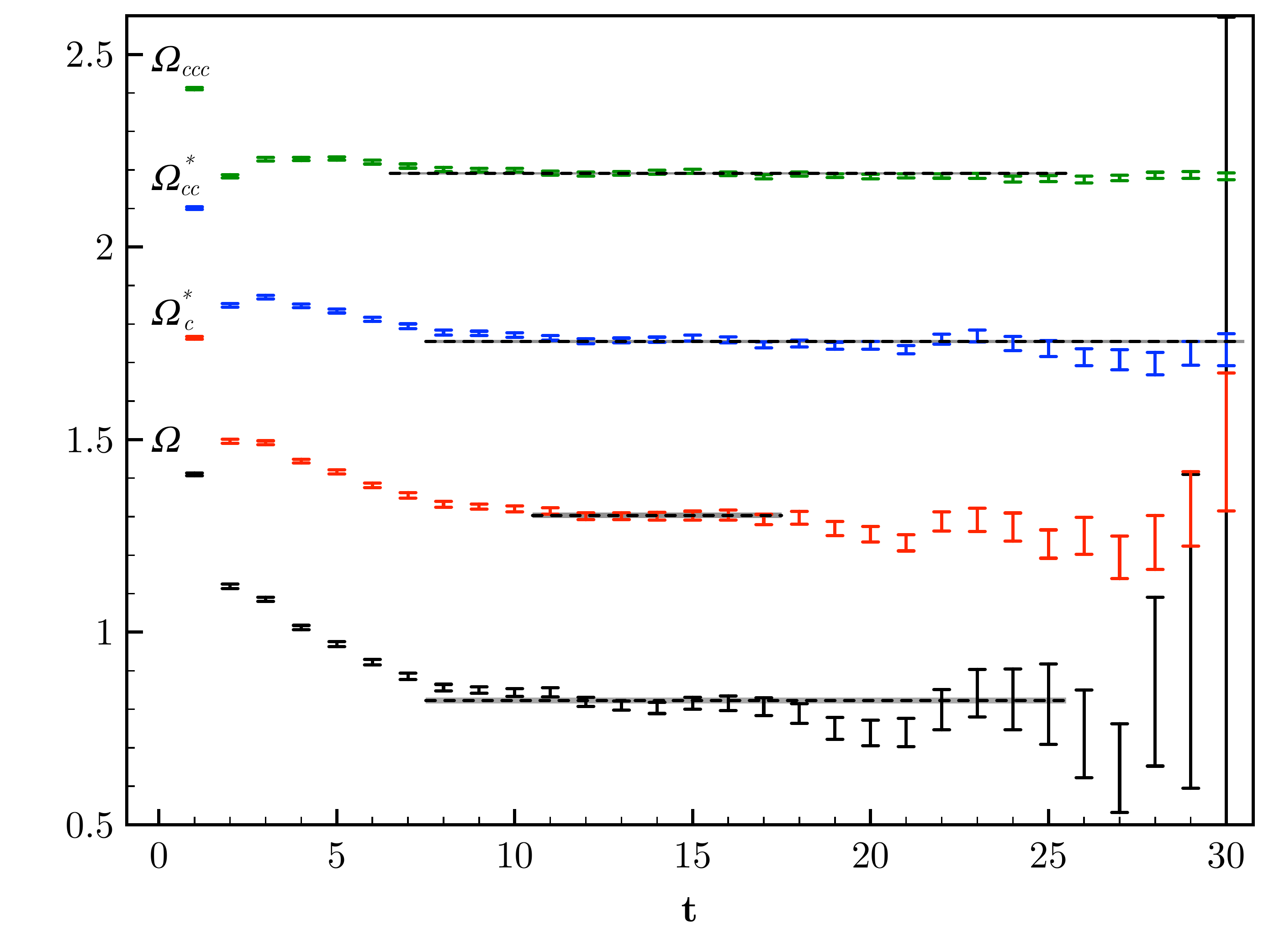}
	\caption{ Effective mass plots for the spin-3/2 baryons. Dashed horizontal lines indicate the fit regions.}
	\label{Fig:effmass}
\end{figure}

The masses of the $\Omega$, $\Omega_c^{(\ast)}$, $\Omega_{cc}^{(\ast)}$ and $\Omega_{ccc}$ baryons are extracted from the shell-source/point-sink lattice two-point function given in Eq.\eqref{deltacf} by a simultaneous fit to all spatial Lorentz components. We show the effective mass plots for the spin-3/2 baryons in Fig~\ref{Fig:effmass}. Results are given in Table~\ref{bar_mass} 
along with a comparison to the masses reported by PDG and other lattice collaborations. 

We identify the lower end, $t_{min}$, of the fit regions $[t_{min}, t_{max}]$ by searching for a time slice where a plateau forms in the effective mass plots. The upper end, $t_{max}$, extends up to time slices where the signal starts to deteriorate. We note that shifting $t_{min}$ to larger time slices gives lower mass values that are closer to those of other lattice groups or the experimental ones. However, the effective mass plots are known to exhibit a fluctuation after a plateau is formed~\cite{Aoki:1995bb} and shifting the fit region would introduce a bias to the determination of the masses.

As compared to the experimentally available results there is a discrepancy of around $100$ MeV in the case of $\Omega$ and $\Omega_c^{(*)}$ masses. Note that the differences may arise from our choice of the strange and charm quark hopping parameters. In order to avoid the partial-quenching effects we have chosen $\kappa_s$ to be the same as that of the sea quark. On the other hand, our $\Omega$ mass is in good agreement with the mass reported by the PACS-CS Collaboration~\cite{Aoki:2008sm}. A retuning of $\kappa_s$ so as to obtain the physical $K$ mass would be desirable for precision calculations. However we expect such a retuning to have a minimal effect on the conclusions of this work. In Ref.~\cite{Can:2013zpa}, by re-tuning $\kappa_c$ mildly, we have indeed confirmed that a variation of baryon masses around $\pm 100$~MeV has minimal effect on the form factors. We have found that such a tuning changes the charge radii by less than $2\%$.  

The mass of the $\Omega_{ccc}$ as obtained in our simulations agrees with those from other lattice simulations~\cite{Namekawa:2013vu,Alexandrou:2014sha,Briceno:2012wt,Brown:2014ena} with different actions. This can be taken as a good indicator for the aptness of the charm-quark action we employ. 
 
\begin{table*}[ht]
	\caption{ $\Omega$, $\Omega_c~(\frac{1}{2}^+, \frac{3}{2}^+)$, $\Omega_{cc}~(\frac{1}{2}^+, \frac{3}{2}^+)$ and $\Omega_{ccc}$ masses (at a pion mass of $m_\pi=156$~MeV) together with the experimental values~\cite{Agashe:2014kda} and those obtained by PACS-CS~\cite{Namekawa:2013vu} (at the physical point, except  $\Omega$ which is at $m_\pi=156$~MeV~\cite{Aoki:2008sm}). We have also included results by ETMC~\cite{Alexandrou:2014sha}, Briceno \emph{et al.}~\cite{Briceno:2012wt} and Brown \emph{et al.}~\cite{Brown:2014ena}. All values are given in units of GeV.
}
\begin{center}
	{
	\setlength{\extrarowheight}{7pt}
\begin{tabular*}{\textwidth}{@{\extracolsep{\fill}}l|ccccccc}
			\hline\hline 
			  & $J^P$ & This work & PACS-CS~\cite{Namekawa:2013vu} &  ETMC~\cite{Alexandrou:2014sha} & Briceno \emph{et al.}~\cite{Briceno:2012wt} & Brown \emph{et al.}~\cite{Brown:2014ena}& Exp.~\cite{Agashe:2014kda}   \\
			\hline \hline
													& &[GeV] & [GeV] & [GeV] & [GeV] & [GeV] \\
			$\Omega_c$ 					& $\frac{1}{2}^+$ & 2.783(13) & 2.673(17) & 2.629(22) & 2.681(48) & 2.679(57) & 2.695(2)\\
			$\Omega_{cc}$ 			& $\frac{1}{2}^+$ & 3.747(10) & 3.704(21) & 3.654(18) & 3.679(62) & 3.738(40) & ---\\
			\hline						
			$\Omega$						& $\frac{3}{2}^+$ & 1.790(17) & 1.772(7)~\cite{Aoki:2008sm} & 1.672(18) & --- & --- & 1.673(29)\\	
			$\Omega_c^\ast$ 		& $\frac{3}{2}^+$ & 2.837(18) & 2.738(17) & 2.709(26) & 2.764(49) & 2.755(61) & 2.766(2)\\
			$\Omega_{cc}^\ast$ 	& $\frac{3}{2}^+$ & 3.819(10) & 3.779(23) & 3.724(21) & 3.765(65) & 3.822(42) & --- \\			
			$\Omega_{ccc}$			& $\frac{3}{2}^+$ & 4.769(6) & 4.789(27) & 4.733(18) & 4.761(79) & 4.796(26) & --- \\
			\hline \hline			
\end{tabular*}
	\label{bar_mass}
	}
\end{center}
\end{table*}

\begin{figure}[ht]
	\centering
	\includegraphics[width=0.475\textwidth]{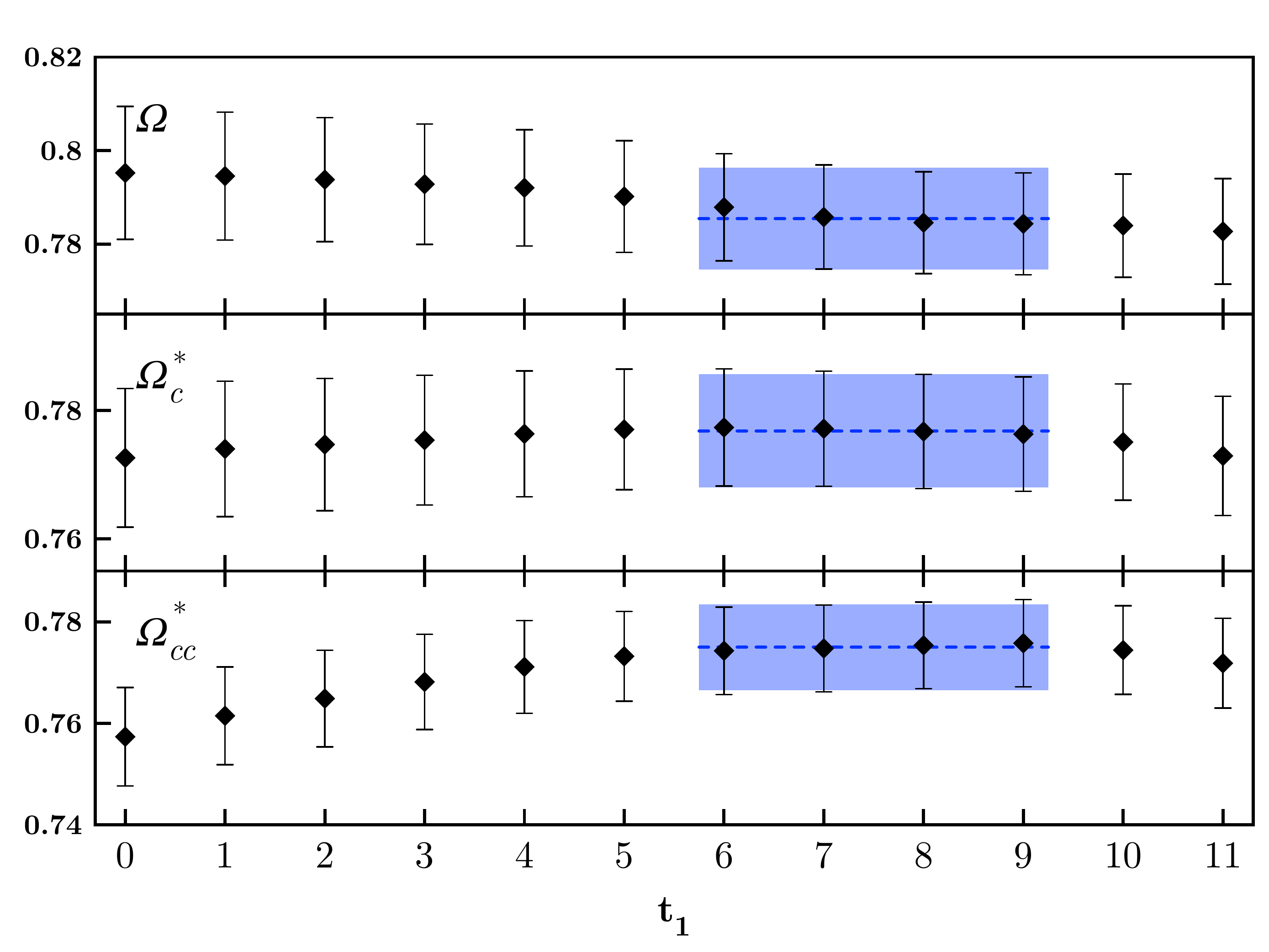}
	\includegraphics[width=0.475\textwidth]{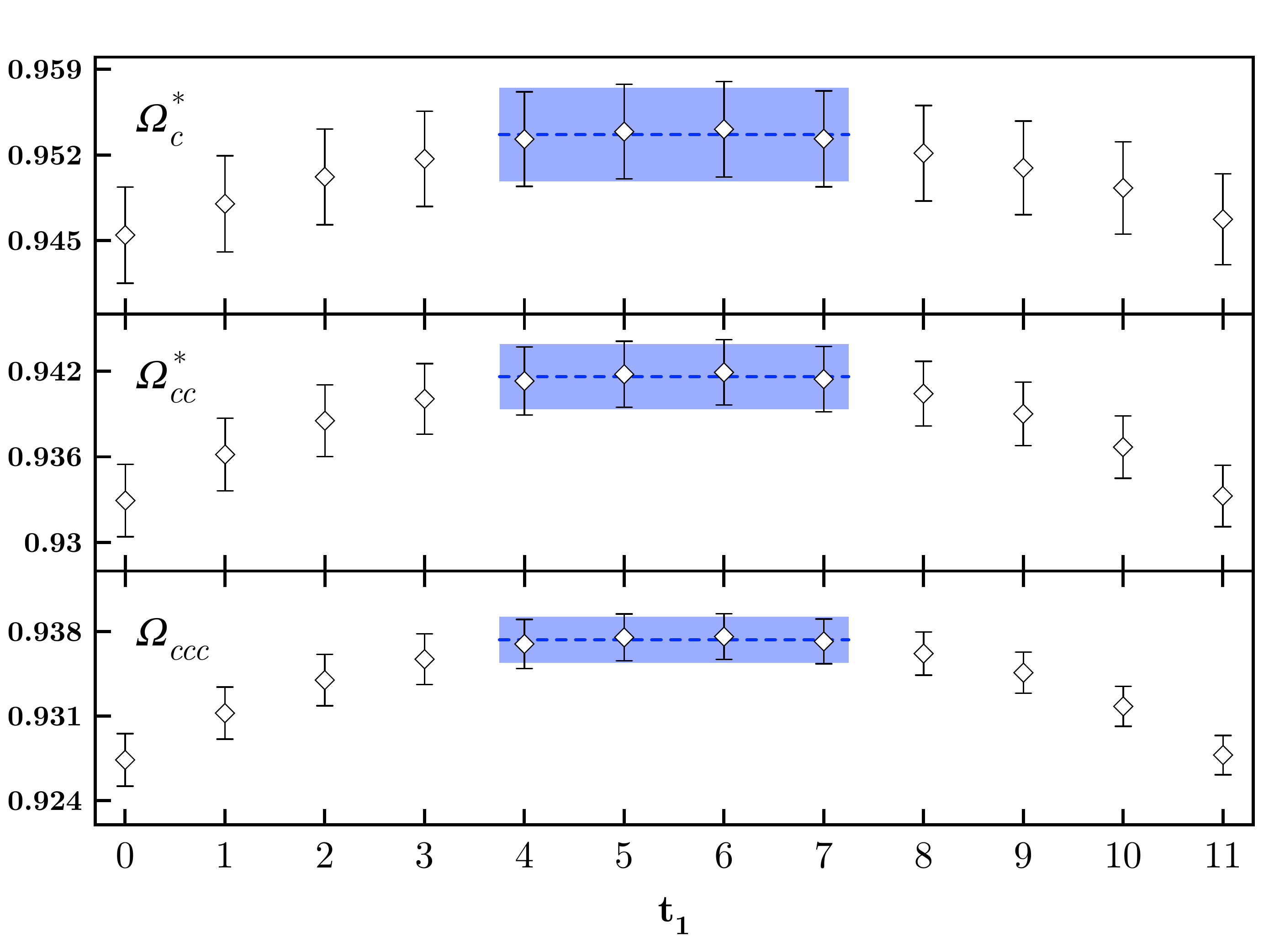}
	\caption{ Strange (filled) and charm-quark (empty) contributions to the $E0$ form factor at the lowest allowed three-momentum transfer ($\mathbf{q}^2$=0.183 GeV$^2$). The contributions are shown for single quark and normalised to unit charge. The fit regions are $t_1=[4,7]$ for the charm sector and $t_1=[6,9]$ for the strange sector.}
	\label{Fig:E0_plato}
\end{figure}	
\begin{figure}[ht]
	\centering
	\includegraphics[width=0.475\textwidth]{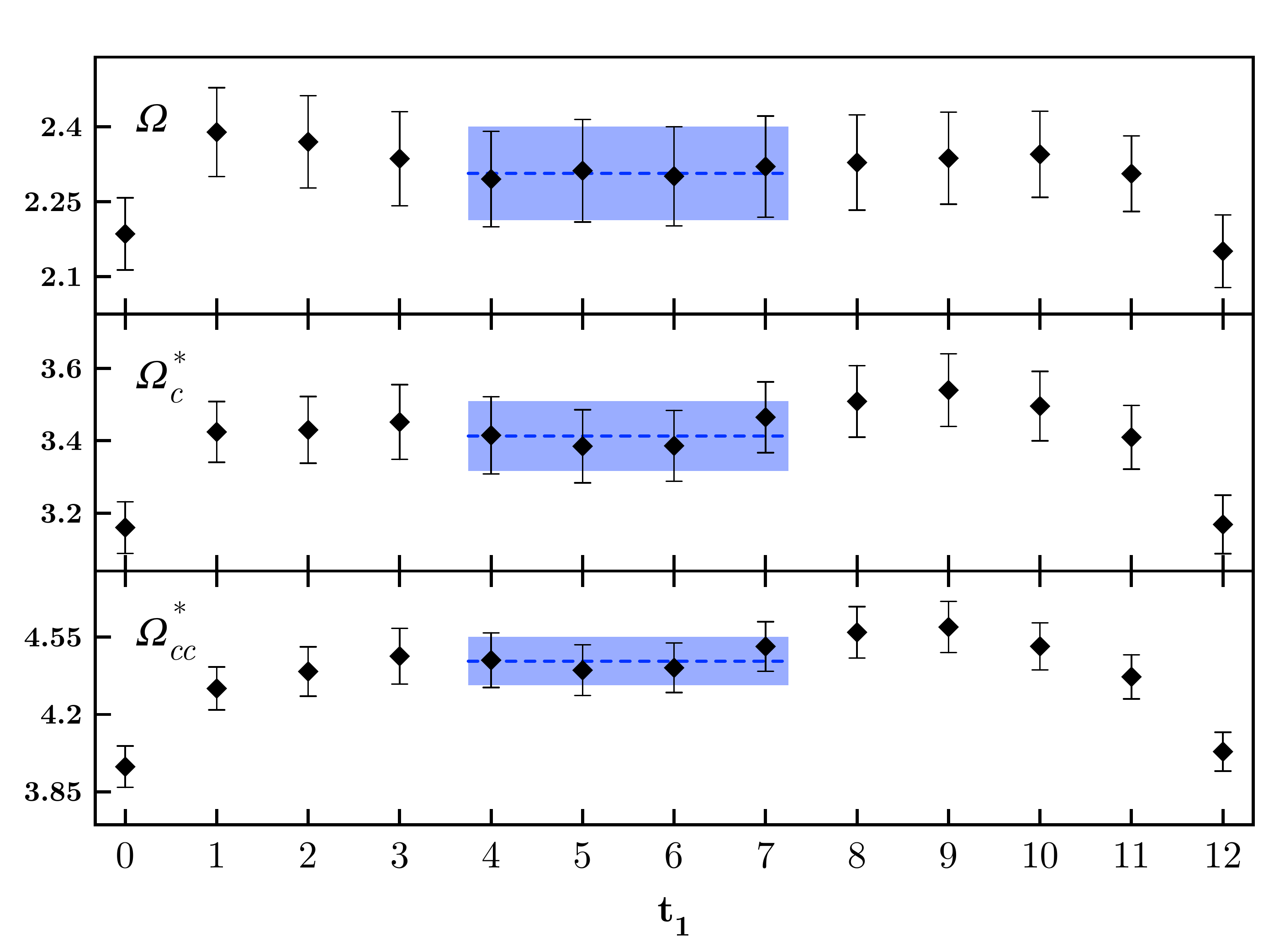}
	\includegraphics[width=0.475\textwidth]{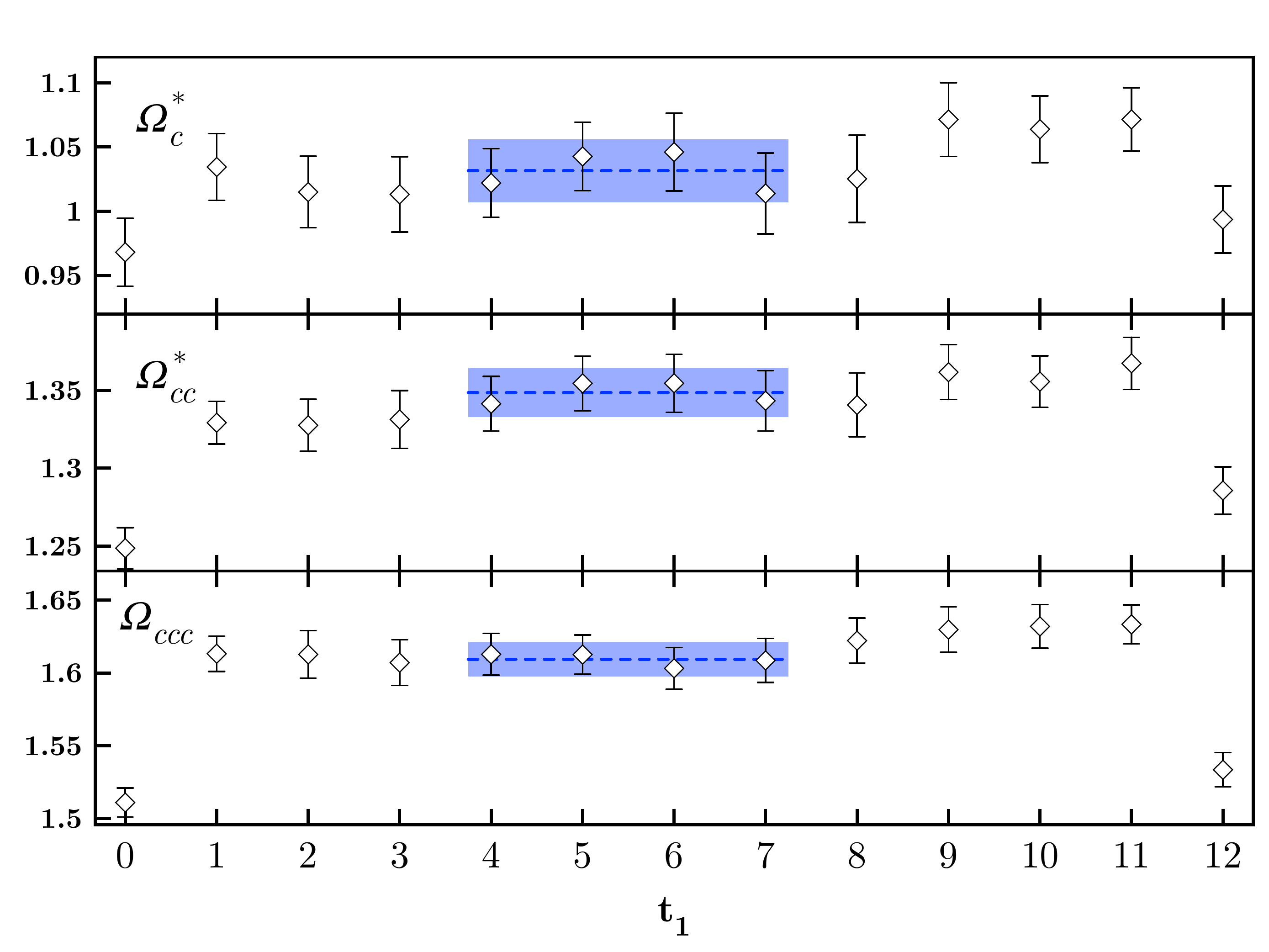}
	\caption{ Same as Fig.~\ref{Fig:E0_plato} but for the $M1$ form factor. Fit region is $t_1 = [4,7]$ for all cases.}
	\label{Fig:M1_plato}
\end{figure}	
\begin{figure}[ht]
	\centering
	\includegraphics[width=0.475\textwidth]{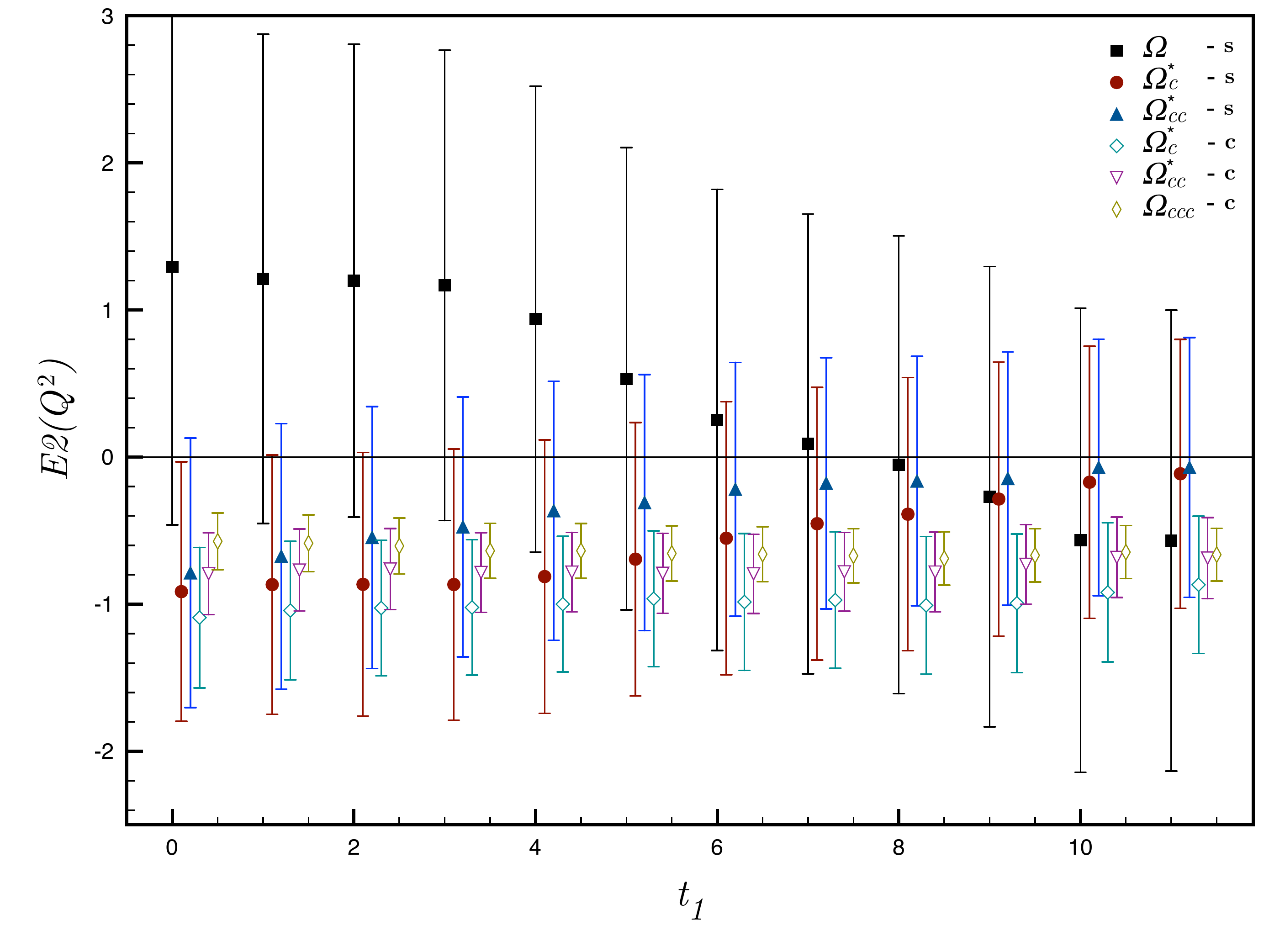}
	\caption{Same as Fig.~\ref{Fig:E0_plato} but for the $E2$ form factor. Fit region is $t_1 = [4,7]$ for all cases. Error bars are slightly shifted for clear view.}
	\label{Fig:E2_plato}
\end{figure}	
\subsection{Form Factor correlation functions}
In our simulations we evaluate each quark sector separately and normalise to unit charge contributions. The baryon properties are then estimated by combining the quark contributions with their weights from respective quark numbers and electric charges as,
\begin{equation}
	\label{obs_comb}
	\langle \mathcal{O} \rangle = N_s e_s \langle \mathcal{O}_s \rangle + N_c e_c \langle \mathcal{O}_c \rangle,
\end{equation} 
where $\langle \mathcal{O} \rangle$ is the observable, $N_q$ is the number quarks inside the baryon having  flavor $q$ and $e_q$ is the electric charge of the quark.

We have extracted the spin-3/2 baryon multipole form factor values by searching for plateau regions of the ratio given in Eq.\eqref{plato_ratio}. The correlation-function ratios for the $E0$, $M1$ and $E2$ form factors are depicted in Figs.~\ref{Fig:E0_plato}-\ref{Fig:E2_plato}. Fit values for the form factors at the lowest allowed three-momentum transfer ($\mathbf{q}^2$=0.183 GeV$^2$) are given in Table~\ref{FFQ2_res_table}. Note that $E0$ form factor reduces to the electric charge of the baryon as usual and the other form factors cannot be directly obtained at zero momentum transfer due to their definitions in Eqs.~\eqref{E2lat}-\eqref{M3lat}. 

\begin{table*}[ht]
	\caption{The values of $E0(Q^2)$, $M1(Q^2)$ and $E2(Q^2)$ form factors at $\mathbf{q}^2$=0.183 GeV$^2$ for  $\Omega$, $\Omega_c^\ast$, $\Omega_{cc}^\ast$ and $\Omega_{ccc}$. Results are given in lattice units for single quark and normalised to unit charge.}
	\label{FFQ2_res_table}
\begin{tabularx}{\textwidth}{l|CC|CC|CC}
		\hline\hline
													& $E0^s(Q^2)$ & $E0^c(Q^2)$ & $M1^s(Q^2)$ & $M1^c(Q^2)$ & $E2^s(Q^2)$ & $E2^c(Q^2)$ 	\\
		\hline\hline
		$\Omega$  		    		& 0.789(12) 	& ---	      	& 2.307(94) 	& --- 				& -0.228(773) & --- 					\\
		$\Omega_c^\ast$  	    		& 0.778(9) 		& 0.954(4) 		& 3.413(96) 	& 1.032(25) 	& -0.630(915) & -0.979(456)  		\\
		$\Omega_{cc}^\ast$  				& 0.775(8) 		& 0.942(2)		& 4.442(110)  & 1.349(16) 	& -0.280(852) & -0.787(266)  	\\
		$\Omega_{ccc}$ 		 		& ---	      	& 0.937(2) 		& ---   			& 1.609(12) 	& --- 				& -0.655(182)   \\
		\hline\hline		
	\end{tabularx}
\end{table*}

\subsection{Charge radii}
\label{Sec:ecr}
Electric charge radius of the baryons are obtained by calculating the slope of the $E0$ form factor at zero-momentum transfer:
\begin{equation}
	\label{chrgR}
	\langle r^2_{E} \rangle = - 6 \frac{d}{dQ^2} G_{E0}(Q^2) |_{Q^2=0}.
\end{equation}
In the case of the proton, the low-Q$^2$ experimental data is well-described by the dipole form Ansatz
\begin{equation}
	\label{dipole}
	G_{E0}(Q^2) = \frac{G_{E0}(0)}{(1 + Q^2/\Lambda^2)^2},
\end{equation} 
where $\Lambda$ is the dipole mass. We assume that such Ansatz also holds for the baryons we study here. Since we perform our simulations with a single value of the finite momentum transfer, a dipole fit of the form factor to a momentum region is not possible. We can, however, extract the charge radii using the expression 
\begin{equation}
	\label{chrgR_dipole}
		\frac{\langle r^2_{E} \rangle}{G_{E0}(0)} = \frac{12}{Q^2_{min}} \left( \sqrt{\frac{G_{E0}(0)}{G_{E0}(Q^2_{min})}} - 1\right),
\end{equation}
which can be readily derived by inserting Eq.\eqref{dipole} into Eq.\eqref{chrgR}.
Our numerical values for the electric charge radii are given in Table~\ref{E_res_table}. Note that the quark sector contributions are for individual quarks of unit electric charge.

\begin{figure}[ht]
	\centering
	\includegraphics[width=0.45\textwidth]{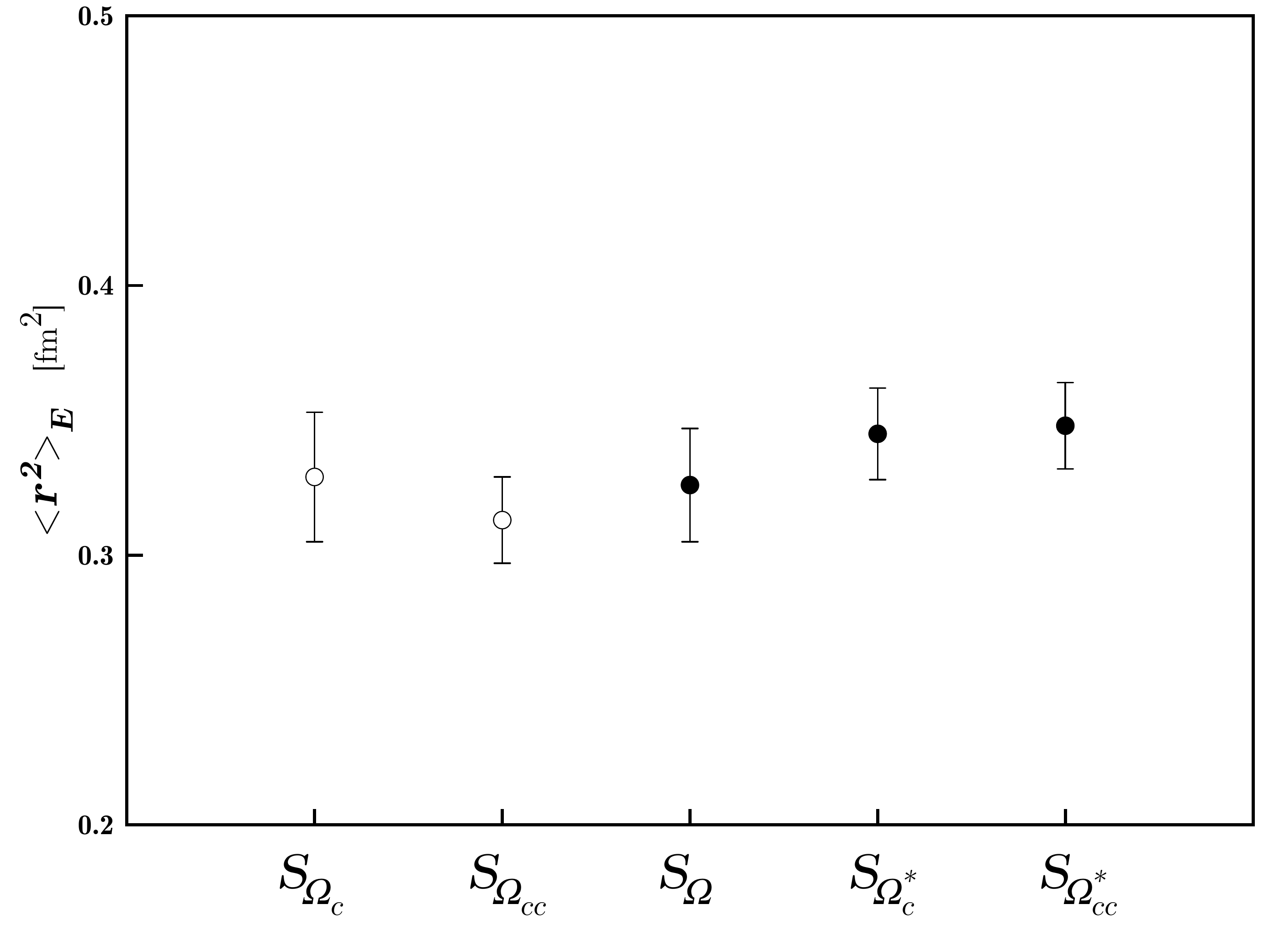}
	\caption{$s$-quark contribution to the electric charge radii of the spin-1/2 $\Omega_c$, $\Omega_{cc}$ and spin-3/2 $\Omega$, $\Omega^\ast_c$ and $\Omega^\ast_{cc}$ baryons.}
	\label{Fig:r2es}
\end{figure}

\begin{figure}[ht]
	\centering
	\includegraphics[width=0.45\textwidth]{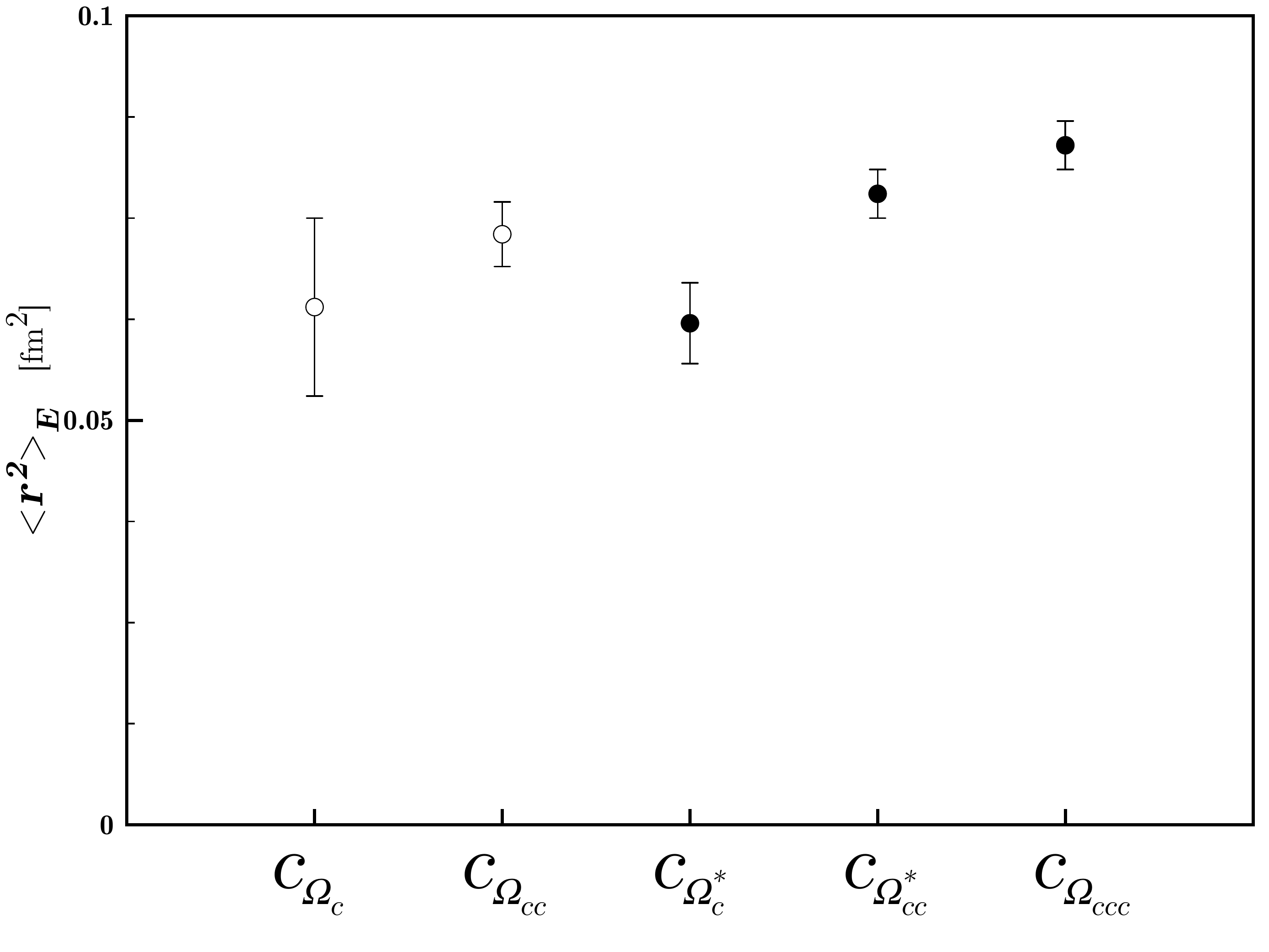}
	\caption{$c$-quark contribution to the electric charge radii of the spin-1/2 $\Omega_c$, $\Omega_{cc}$ and spin-3/2 $\Omega^\ast_c$, $\Omega^\ast_{cc}$ and $\Omega_{ccc}$ baryons.}
	\label{Fig:r2ec}
\end{figure}

\begin{table}[ht]
	\caption{Electric charge radii of the $\Omega$, $\Omega_c^\ast$, $\Omega_{cc}^\ast$ and $\Omega_{ccc}$. Results are given in fm$^2$. Quark sector contributions are for single quark and normalised to unit charge. Electric charge radii of spin-1/2 baryons are estimated through form factor fits as in Ref.~\cite{Can:2013tna}. Total electric charge radius of the spin-1/2 $\Omega_c$ is estimated by the Eq.~\ref{obs_comb} since its electric form factor vanishes due to its zero electric charge.}
	\label{E_res_table}
	\begin{tabularx}{0.5\textwidth}{l|CCC}
		\hline\hline
													& $\langle r_E^2 \rangle_s$ & $\langle r_E^2 \rangle_c$ & $\langle r_E^2 \rangle$ \\
		\hline\hline		
													&  [fm$^2$]      						& [fm$^2$]      						& [fm$^2$] 								\\		
		$\Omega_c$						& 0.329(25) 								& 0.064(11) 								& -0.177(18)							\\
		$\Omega_{cc}$					& 0.313(16) 								& 0.073(4) 									& 0.026(4)								\\		
		\hline
		$\Omega$ 			 		    & 0.326(21) 								& ---	      								& -0.326(21) 							\\
		$\Omega_c^\ast$ 			& 0.345(17) 								& 0.062(5) 									& -0.189(12) 							\\
		$\Omega_{cc}^\ast$  	& 0.348(16) 								& 0.078(3)									& -0.012(6)  							\\
		$\Omega_{ccc}$ 			 	& ---	     	 								& 0.084(3) 									& 0.168(5)   							\\
		\hline\hline		
	\end{tabularx}
\end{table}
We observe that the $s$-quark contribution to the electric charge radii is almost independent of the quark-flavor composition of the baryon. The charge radii of both spin-1/2 and spin-3/2 baryons agree within one standard deviation, which can be seen more clearly in Fig.~\ref{Fig:r2es}. In the case of $c$-quark contributions illustrated in Fig.~\ref{Fig:r2ec}, the charge radii in all baryons are similar. Although the contribution of individual $c$-quark seems to increase slightly as the number of $c$-quark increases in the composition of the baryon, this change remains negligibly small. 

\begin{figure}[ht]
	\centering
	\includegraphics[width=0.45\textwidth]{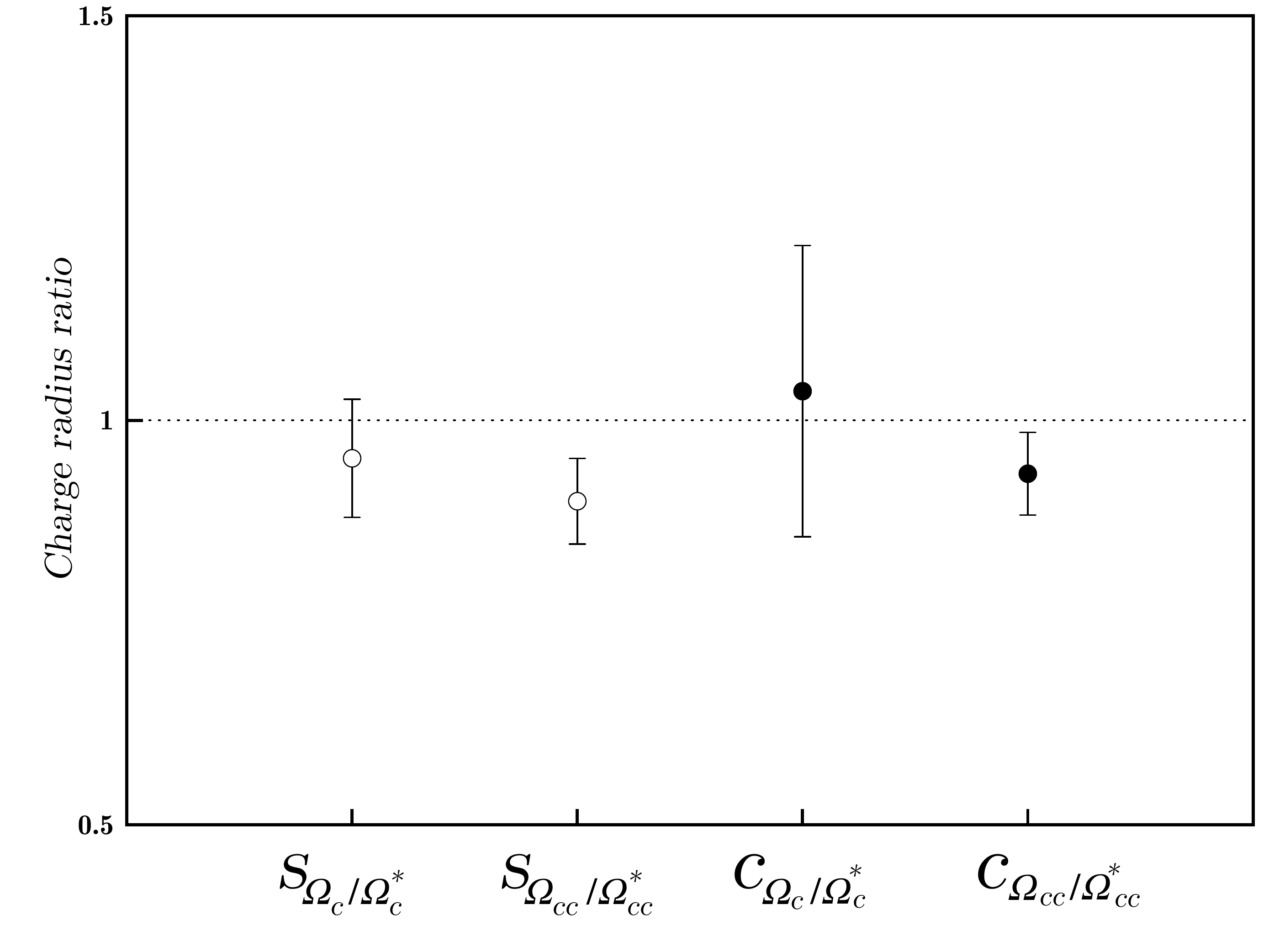}
	\caption{Ratio of the quark contribution to the electric charge radii of the spin-1/2 $\Omega_c$, $\Omega_{cc}$ to spin-3/2 $\Omega$, $\Omega^\ast_c$ baryons.  $q_{B/B^\ast}$ is a shorthand notation for the ratio, $\left< r^2_E\right>^q_B / \left< r^2_E\right>^q_{B^\ast}$, where $q$ is the quark flavor and $B$ is the baryon.}
	\label{Fig:r2esc}
\end{figure}

In Fig.~\ref{Fig:r2esc} we show the ratios of individual quark-flavor contributions in the spin-1/2 to that in the spin-3/2 sector. We observe that for the singly-charmed $\Omega_c$ baryon, $s$- and $c$-quark charge distributions are insensitive to the spin-flip of the $c$-quark whereas in the case of the doubly-charmed $\Omega_{cc}$ baryon the contributions of $s$- and $c$-quark to the charge radii increase. 

 
\begin{figure}[ht]
	\centering
	\includegraphics[width=0.45\textwidth]{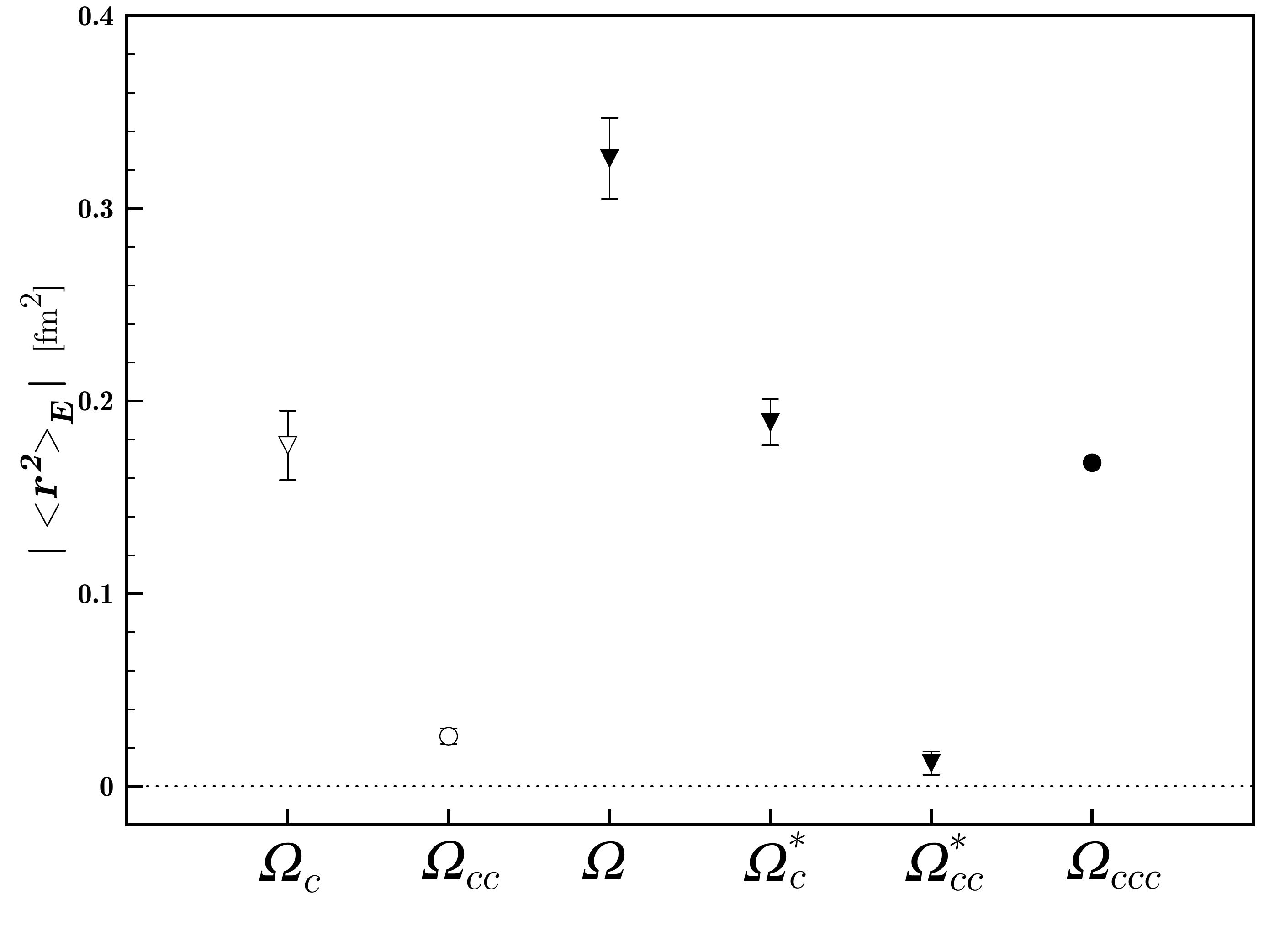}
	\caption{The electric charge radii of the spin-1/2 $\Omega_c$, $\Omega_{cc}$ and spin-3/2 $\Omega$, $\Omega^\ast_c$ and $\Omega^\ast_{cc}$ baryons. Absolute values are shown for a better comparison. Data points denoted by a triangle indicate a negative value.}
	\label{Fig:r2e12-32}
\end{figure}

We combine the individual quark contributions according to Eq.~\eqref{obs_comb} and list the numerical results in Table~\ref{E_res_table}. We find the electric charge radius of the $\Omega$ baryon to be $\langle r_E^2 \rangle_{\Omega^-} = -0.326(21)$ fm$^2$ in quite good agreement with the previous lattice determinations \cite{Alexandrou:2010jv,PhysRevD.80.054505}. A comparison of baryon charge radii is made in Fig.~\ref{Fig:r2e12-32}. In magnitude, $\Omega$ baryon has the largest electric charge radius among all baryons we study. Spin-1/2 (spin-3/2) $\Omega_c$ ($\Omega_c^\ast$) and $\Omega_{ccc}$ seem to have similar charge radii while the $\Omega_{cc}$ ($\Omega_{cc}^\ast$) baryon has almost a vanishing charge radius.

Based on the similarity of the quark contributions to the charge radii one can naively assume that the quark sector contributions to the charge radii to be similar for all spin-3/2 baryons that we consider so that, $\langle {r_E^2} \rangle^s_\Omega = \langle r_E^2 \rangle^s_{\Omega_c^\ast} = \langle r_E^2 \rangle^s_{\Omega_{cc}^\ast} = R^2_s$ and $\langle r_E^2 \rangle^c_{\Omega_c^\ast} = \langle r_E^2 \rangle^c_{\Omega_{cc}^\ast} = \langle r_E^2 \rangle^c_{\Omega_{ccc}} = R^2_c$. Using  Eq.~\eqref{obs_comb} we can derive a relation between the electric charge radii of the spin-3/2 baryons as, $\left( \langle {r_E^2} \rangle_{\Omega_c^\ast} + \langle {r_E^2} \rangle_{\Omega_{ccc}} \right)/2 = \langle {r_E^2} \rangle_{\Omega_{cc}^\ast}$. Comparing $\left(\langle {r_E^2} \rangle_{\Omega_c^\ast} + \langle {r_E^2} \rangle_{\Omega_{ccc}} \right)/2 = -0.011(8)$, as obtained from such an estimation and the computed charge radius of $\Omega_{cc}^\ast$, $\langle {r_E^2} \rangle_{\Omega_{cc}^\ast} = -0.012(6)$, this relation seems to hold nicely. This implies that the individual quark contributions for each flavor to charge radii are similar for all baryons we consider here and their radii differ due to different quark compositions they have.    

It may be instructive to compare the contributions of the strange quark to the electric charge radii in the case of $\Omega_c^\ast$(ssc), $\Omega_{cc}^\ast$(scc) and $\Xi^\ast$(ssu), $\Sigma^\ast$(suu) baryons. The latter have been calculated in Ref.~\cite{PhysRevD.80.054505}. Such a comparison would provide a better understanding as to how the charge radii are affected when the light quark is exchanged by a charm quark. A comparison of $s$-quark electric charge radii in $\Omega_c^\ast$ - $\Xi^\ast$ in Table~\ref{s_comp_table} reveals  the effect of changing the single $u$-quark by a $c$-quark: When the singly represented quark is heavier, the $s$-quark charge radius increases. In case of the $\Omega_{cc}^\ast$ - $\Sigma^\ast$ baryons, the doubly represented light quarks are changed to $c$-quarks. While the current precision does not allow a clear conclusion, such a comparison again suggests an increase in the charge radius.
              
\begin{table}[t]
	\caption{Single strange quark contribution normalized to the electric charge radii of the $\Omega_c^\ast$ and $\Omega_{cc}^\ast$ (normalized to unit charge) in comparison to that of the decuplet $\Xi^\ast$ and $\Sigma^\ast$. Decuplet values are taken from tables XI. and XII. of Ref.~\cite{PhysRevD.80.054505}.}
	\label{s_comp_table}
	\begin{tabularx}{0.5\textwidth}{l|CC|CC}
		\hline\hline
																				& $S_{\Omega_c^\ast}$ & $S_{\Xi^\ast}$ 	& $S_{\Omega_{cc}^\ast}$ 	& $S_{\Sigma^\ast}$ \\
		\hline\hline		
		$\langle r_E^2 \rangle$	[fm$^2$]	& 0.345(17)				& 0.308(17)		& 0.348(16)						& 0.321(22)			\\
		\hline\hline		
	\end{tabularx}
\end{table}
\subsubsection*{Findings for the electric charge radii}
\begin{itemize}
	\item Strange quark charge radii are insensitive to the baryon quark-flavor composition. \\
	\item Charm-quark charge radii increase as the number of charm quarks increases in the composition of the baryon. \\
	\item For singly charmed baryons, $s$- and $c$-quark charge radii are not affected by the spin-flip whereas the charge radii of doubly charmed baryons increase. \\
	\item We find the electric charge radius of $\Omega^-$ to be $\langle r_E^2 \rangle_{\Omega^-} = -0.326(21)$ fm$^2$. \\
\end{itemize}

\subsection{Magnetic moments}
\label{Sec:mmoms}
Magnetic moments of the baryons are related to the $Q^2=0$ value of the magnetic form factor $M1$. We evaluate the magnetic dipole moment in units of nuclear magnetons,
\begin{equation}
	\label{magmom}
	\mu_{B}=G_{M1}(0)\frac{e}{2m_{B}} = G_{M1}(0)\frac{m_N}{m_{B}}\mu_N,
\end{equation}
where $m_N$ is the physical nucleon mass and $m_{B}$ is the baryon mass as obtained on the lattice.

In order to make contact with the value of the magnetic form factor at zero momentum transfer, $G_{M1}(0)$, we apply the procedure in Ref.~\cite{Leinweber:1992pv} and assume that the momentum-transfer dependence of the multipole form factors is the same as the momentum dependence of the respective baryon's charge form factor. For instance, the scaling of $G_{M1}$ is given by
\begin{equation}\label{scaling}
	G_{M1}^{s,c}(0)=G_{M1}^{s,c}(q^2)\frac{G_{E}^{s,c}(0)}{G_{E}^{s,c}(q^2)},
\end{equation} 
where we consider the scaling of $s$ and $c$ quark sectors separately since each sector has a different scaling property. $G_{M1}(0)$ is then constructed via Eq.\eqref{obs_comb}.

\begin{figure}[ht]
	\centering
	\includegraphics[width=0.45\textwidth]{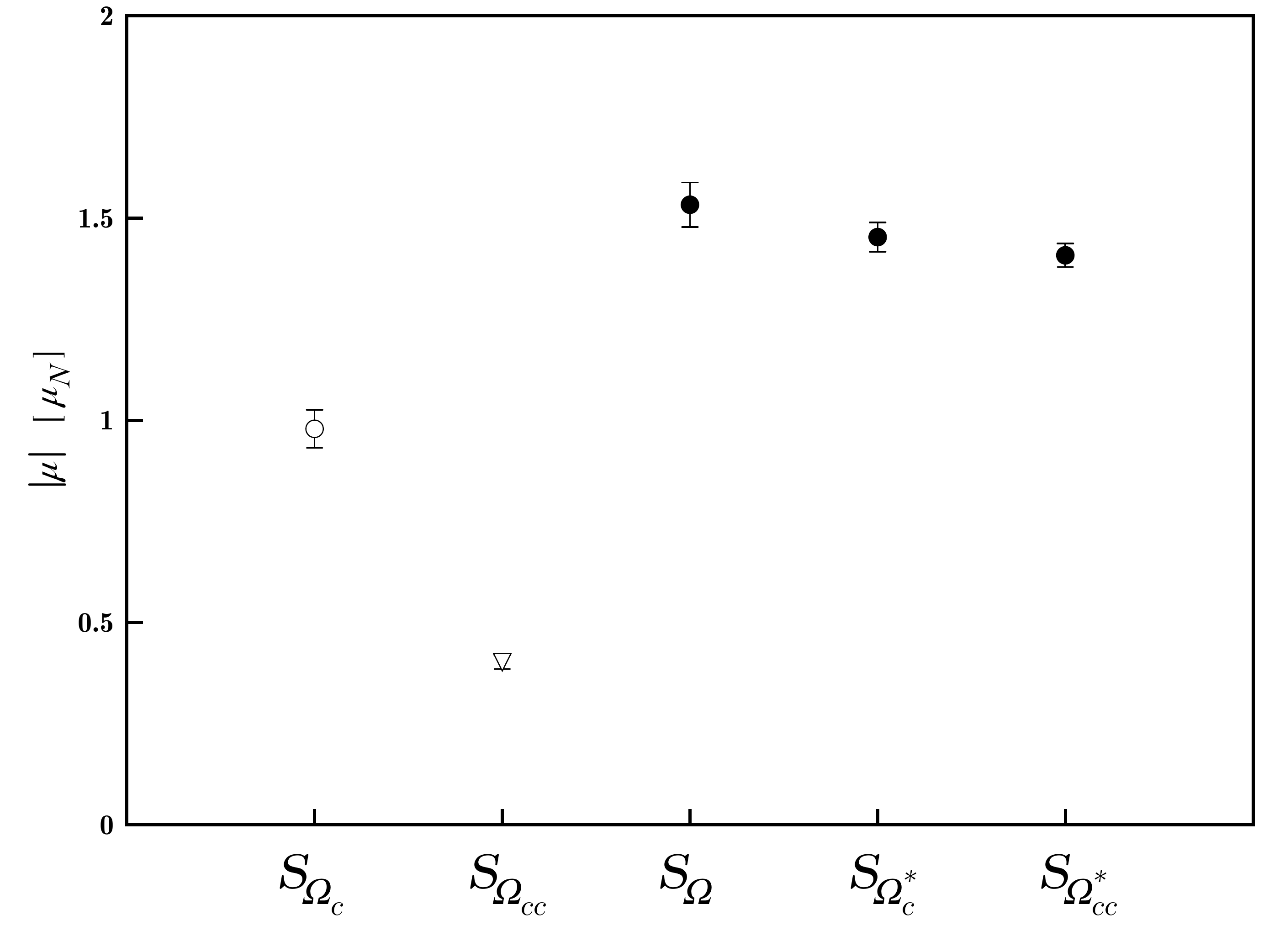}
	\caption{$s$-quark contributions to the magnetic moments of the spin-1/2 $\Omega_c$, $\Omega_{cc}$ and spin-3/2 $\Omega$, $\Omega^\ast_c$ and $\Omega^\ast_{cc}$ baryons. Absolute values are shown for a better comparison. Triangle symbol indicates a negative value.}
	\label{Fig:mus}
\end{figure}

\begin{figure}[ht]
	\centering
	\includegraphics[width=0.45\textwidth]{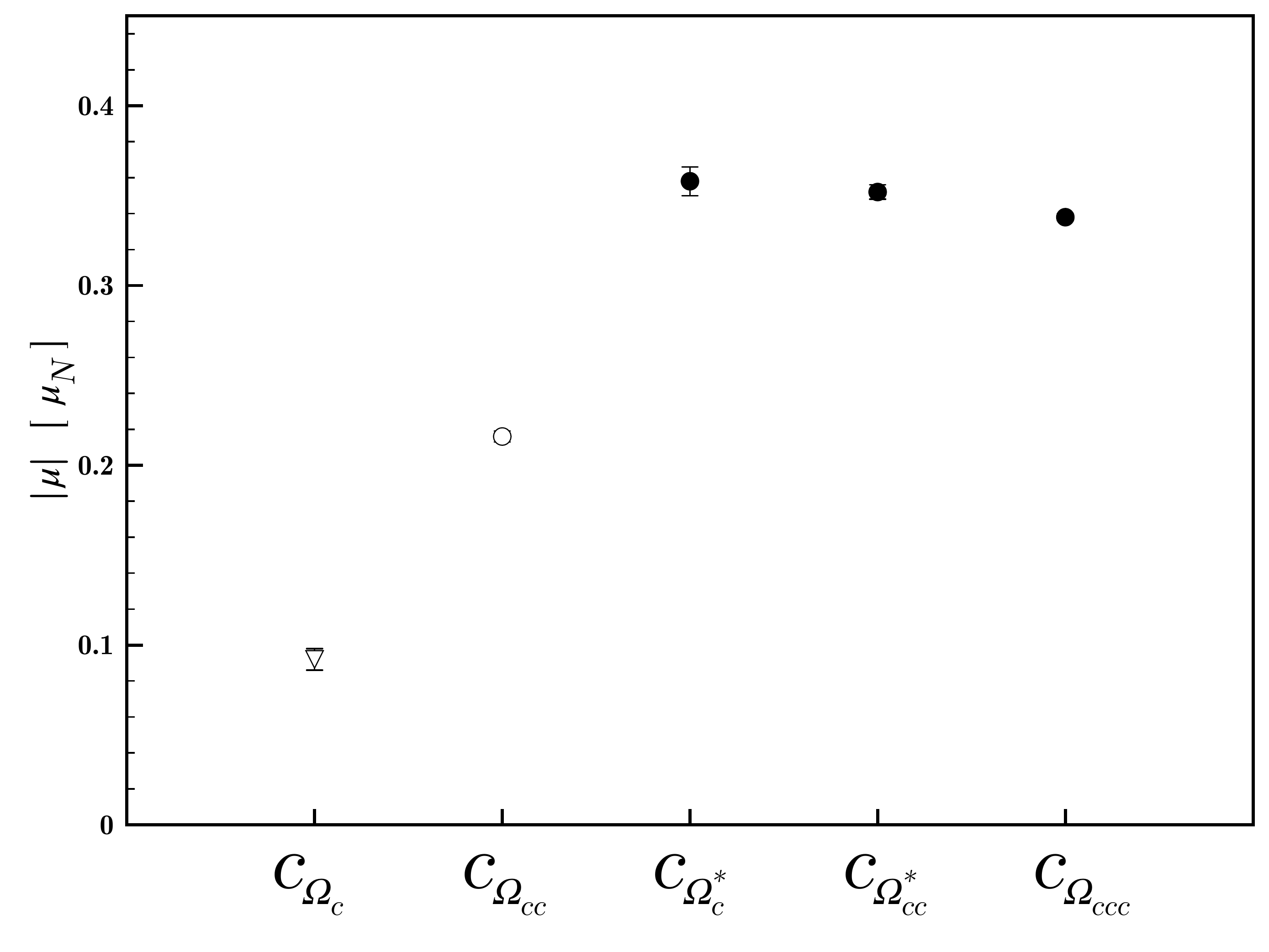}
	\caption{$c$-quark contributions to the magnetic moments of the spin-1/2 $\Omega_c$, $\Omega_{cc}$ and spin-3/2 $\Omega^\ast_c$, $\Omega^\ast_{cc}$ and $\Omega_{ccc}$ baryons. Absolute values are shown for a better comparison. Triangle symbol indicates a negative value.}
	\label{Fig:muc}
\end{figure}

Our numerical values for the magnetic moments are listed in Table~\ref{MM_res_table}, which are also illustrated in Figs.~\ref{Fig:mus} and \ref{Fig:muc}. We find for spin-1/2 baryons that the contribution of a single quark to the magnetic moment significantly increases when it is doubly represented. The doubly-represented quarks make a spin-1 combination according to the Pauli principle and thus the dominant component is aligned with the total spin of the baryon, which leads to such an enhanced contribution compared to the anti-aligned singly-represented sector. A sign change is evident due to the spin flip in case of the $\Omega_c$ and $\Omega_{cc}$ baryons. The $s$-quark contributions in the spin-3/2 sector have a slight tendency to decrease as the number of $s$-quarks decreases. In contrast to this, the $c$-quark contributions tend to decrease as the number of $c$-quarks increases. 

A comparison of the light sector to the heavy sector may reveal the differences between the systems having the same spin but different composition of quark flavors. For instance, we may compare the $s$-quark magnetic moment in the charmed baryons to that in the light decuplet baryons, namely the $\Omega_c^\ast$(ssc) and $\Xi^\ast$(ssu) baryons or the $\Omega_{cc}^\ast$(scc) and $\Sigma^\ast$(suu) baryons. Such a comparison would help us to understand the effects of changing a light quark by a charm quark. Magnetic moments of light decuplet baryons have been calculated in Ref.~\cite{PhysRevD.80.054505} with quenched lattice QCD. In Table~\ref{mm_comp_table_s}, we quote the results obtained with the smallest pion mass, $m_\pi=300$ MeV. As the work in Ref.~\cite{PhysRevD.80.054505} was done on quenched lattices at a much heavier pion mass, a quantitative comparison is not much useful. However, a qualitative comparison can be made as follows: The $s$-quark contributions to the magnetic moments of the charmed and light decuplet baryons are different. The charmed baryons have smaller magnetic moments than the light baryons. 

\begin{table}[t]
	\caption{Magnetic moments of the $\Omega$, $\Omega_c^\ast$, $\Omega_{cc}^\ast$ and $\Omega_{ccc}$. Results are given in units of nuclear magnetons, $\mu_N$. Quark sector contributions are for single quark and normalised to unit charge. Magnetic moments of spin-1/2 baryons are estimated through form factor fits as in Ref.~\cite{Can:2013tna}.}
	\label{MM_res_table}
\begin{tabularx}{0.5\textwidth}{l|CCC}
		\hline\hline
																	& $\mu_s$   	& $\mu_c$   & $\mu$				\\
		\hline\hline
																	& $[\mu_N]$ 	& $[\mu_N]$ & $[\mu_N]$ 	\\		
		$\Omega_c$ 										& 0.979(47) 	& -0.092(6)	& -0.688(31)	\\
		$\Omega_{cc}$									& -0.402(17) 	& 0.216(3)	& 0.403(7)		\\		
		\hline
		$\Omega$ 			 		    				& 1.533(55) 	& ---	   	 	& -1.533(55)  \\
		$\Omega_c^\ast$ 			 	    	& 1.453(36)		& 0.358(8) 	& -0.730(23) 	\\
		$\Omega_{cc}^\ast$ 			 			& 1.408(29) 	& 0.352(4) 	& 0.000(10) 	\\
		$\Omega_{ccc}$ 			 					& ---					& 0.338(2) 	& 0.676(5)		\\
		\hline\hline		
	\end{tabularx}
\end{table}
\begin{table}[t]
	\caption{Strange quark contributions to the magnetic moments of the $\Omega_c^\ast$, $\Omega_{cc}^\ast$, $\Sigma^\ast$ and $\Xi^\ast$. Decuplet baryon results are calculated in Ref.~\cite{PhysRevD.80.054505} on a quenched configuration with $m_\pi = 300$ MeV. All contributions are for a single strange quark of unit charge.}
	\label{mm_comp_table_s}
	\begin{tabularx}{0.5\textwidth}{l|CC|CC}
		\hline\hline
																				& $S_{\Omega_c^\ast}$ & $S_{\Xi^\ast}$ 	& $S_{\Omega_{cc}^\ast}$ 	& $S_{\Sigma^\ast}$ \\
		\hline\hline
		$\mu$	[$\mu_N$]											& 1.453(36)						& 1.725(77)				& 1.408(29)								& 1.750(10)					\\
		\hline\hline
	\end{tabularx}
\end{table}	

\begin{figure}[ht]
	\centering
	\includegraphics[width=0.45\textwidth]{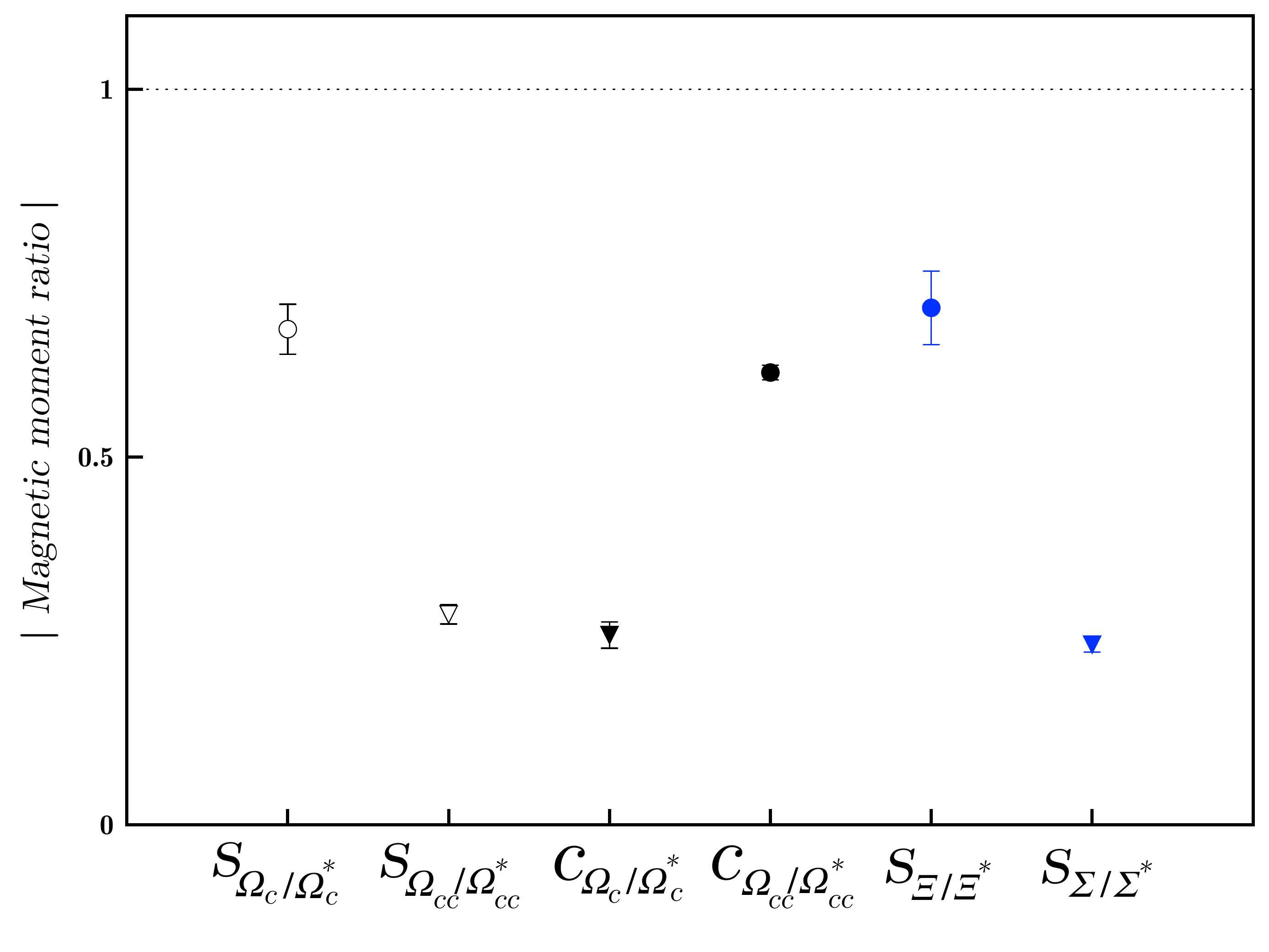}
	\caption{Ratios of the quark contributions to the magnetic moments of the spin-1/2 $\Omega_c$, $\Omega_{cc}$ to spin-3/2 $\Omega$, $\Omega^\ast_c$ baryons. Rightmost blue data points are octet/decuplet ratios calculated using the $m_\pi = 300$ MeV quenched simulation results of Refs.~\cite{PhysRevD.74.093005, PhysRevD.80.054505}. $q_{B/B^\ast}$ is a shorthand notation for $\mu^q_B / \mu^q_{B^\ast}$ where $q$ is the quark flavor and $B$ is the baryon. Absolute values are shown for a better comparison. Data points denoted by a triangle indicate a negative value.}
	\label{Fig:musc}
\end{figure}

Similarly to the electric charge radii in Sec.~\ref{Sec:ecr}, the effect of the quark spin configurations on the quark magnetic moments can be further studied by considering the ratios of the quark-sector contributions to spin-1/2 and spin-3/2 baryons. The numerical values are given in Table~\ref{mm_comp_table_ratio} together with the octet-decuplet ratios extracted from Refs.~\cite{PhysRevD.74.093005, PhysRevD.80.054505}. The ratios are also illustrated in Fig.~\ref{Fig:musc}. It is clearly seen that both the individual $s$-quark and $c$-quark contributions are enhanced in the case of spin-3/2 baryons. The $s$-quark magnetic moment ratio in the case of doubly strange baryons, $S_{\Omega_c^/\Omega_c^\ast}$, and the $c$-quark magnetic moment ratio in doubly charmed baryons, $C_{\Omega_{cc}^/\Omega_{cc}^\ast}$, are consistent with each other. Such a behaviour is also observed in the case of the singly strange and the singly charmed baryon ratios, suggesting that the difference between the spin-1/2 and the spin-3/2 baryons is almost independent of the quark flavour. 

We can further extend the comparison to octet and decuplet sector by including the $\Sigma$, $\Xi$, $\Sigma^\ast$ and $\Xi^\ast$ baryons' strange sector magnetic moments extracted in Refs.~\cite{PhysRevD.74.093005, PhysRevD.80.054505}. For the ease of discussion we compile the numerical values in Table~\ref{mm_comp_table_ratio}. It is seen that the $s$-quark contributions are similar for the doubly strange baryons $S_{\Omega_c / \Omega_c^\ast}$ and $S_{\Xi / \Xi^\ast}$. However in the case of singly-strange baryons, for instance comparing the $S_{\Omega_{cc} / \Omega_{cc}^\ast}$ ratio to $S_{\Sigma / \Sigma^\ast}$ ratio, we observe that the increase in the $s$-quark contribution (as the baryon spin changes from 1/2 to 3/2) is larger when the $s$-quark is accompanied by a \emph{uu} component rather than a \emph{cc} component.  Interestingly, if we make a similar comparison for the singly represented quarks in $C_{\Omega_c/\Omega^\ast_c}$ and $S_{\Sigma/\Sigma^\ast}$, it is seen that the increase in single quark’s contribution to the magnetic moment is less sensitive to the accompanying quark’s flavor. 
\begin{table}[t]
	\caption{Ratios of the quark magnetic moment contributions in $1/2^+$/$3/2^+$. Octet/decuplet ratios are extracted from the numerical results available in the Refs.~\cite{PhysRevD.74.093005, PhysRevD.80.054505}. All values are ratios of a single quark contribution of unit charge.}
	\label{mm_comp_table_ratio}
	\begin{tabularx}{0.5\textwidth}{l|CC|C}		
		\hline\hline
																										&	$S_{\Omega_c / \Omega_c^\ast}$	& $C_{\Omega_{cc} / \Omega_{cc}^\ast}$ & $S_{\Xi / \Xi^\ast}$ \\
		$\left|\mu^q_B / \mu^q_{B^\ast}\right|$					& 0.674(34) & 0.615(10) & 0.703(50) \\
		\hline\hline
																										&	$S_{\Omega_{cc} / \Omega_{cc}^\ast}$	& $C_{\Omega_{c} / \Omega_{c}^\ast}$ & $S_{\Sigma / \Sigma^\ast}$ \\
		$\left|\mu^q_B / \mu^q_{B^\ast}\right|$					& 0.286(13) & 0.258(18) & 0.245(10) \\																				
		\hline\hline																				
	\end{tabularx}
\end{table}

\begin{figure}[ht]
	\centering
	\includegraphics[width=0.45\textwidth]{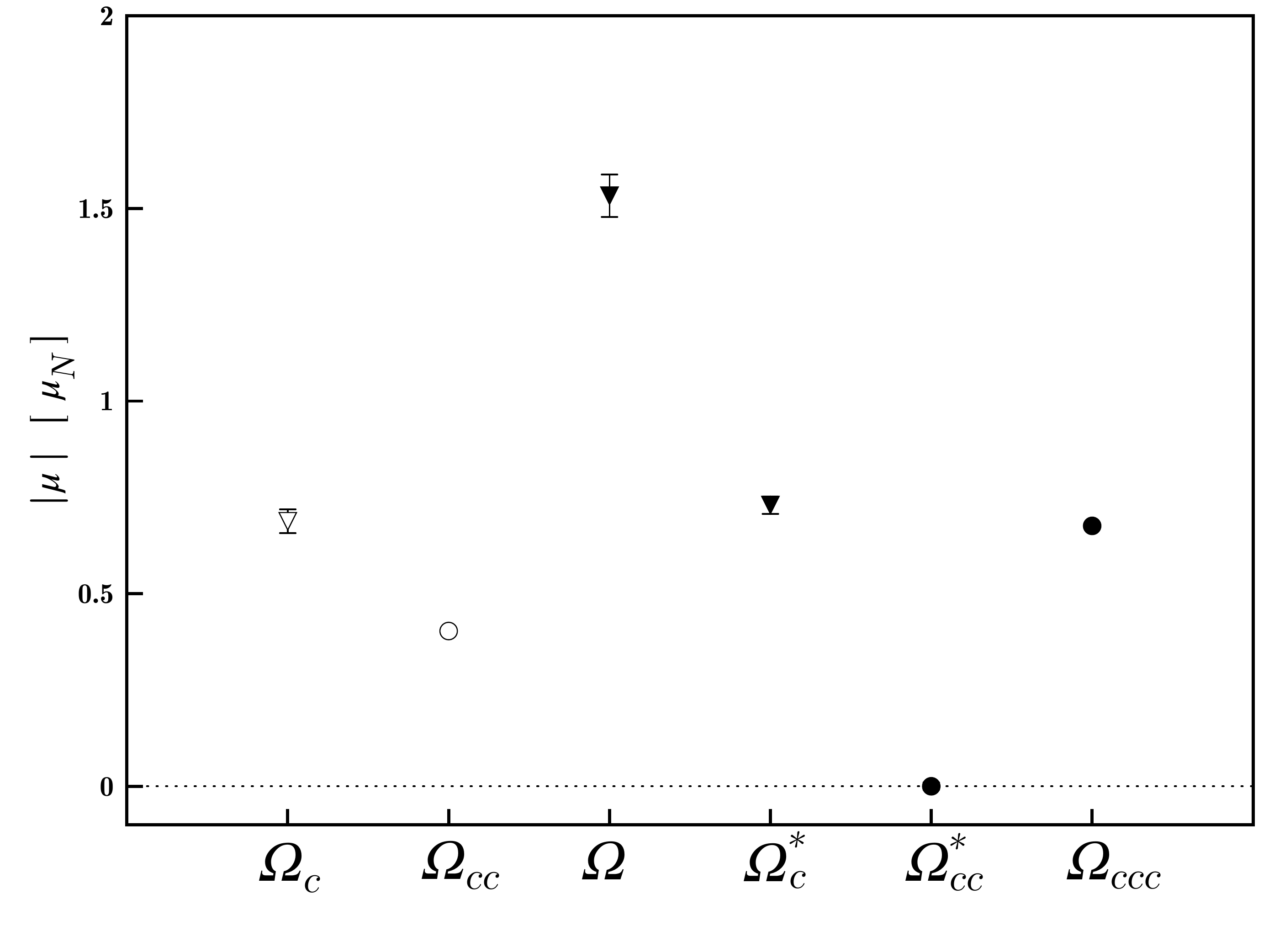}
	\caption{Total magnetic moments of the spin-1/2 $\Omega_c$, $\Omega_{cc}$ and spin-3/2 $\Omega^\ast_c$, $\Omega^\ast_{cc}$ and $\Omega_{ccc}$ baryons. Absolute values are shown for a better comparison. Data points denoted by a triangle indicate a negative value.}
	\label{Fig:mu12-32}
\end{figure}

We calculate the total magnetic moments by combining the quark sectors via Eq.~\eqref{obs_comb}. The numerical values are given in the third column of Table~\ref{MM_res_table} and an illustrative comparison is made in Fig.~\ref{Fig:mu12-32}. We find the magnetic moment of the $\Omega^-$ baryon to be $\mu_{\Omega^-} = -1.533 \pm 0.055$ $\mu_N$, which is smaller in magnitude than the experimental value, $\mu^{exp}_{\Omega^-} = -2.02 \pm 0.05 $ $\mu_N$~\cite{Agashe:2014kda}. Magnetic moments are sensitive to the mass of the baryon. One of the reasons for this discrepancy can arise from the difference between our $m_\Omega = 1.790(17)$ and the experimental value, $m_\Omega = 1.673(29)$, which is around $7\%$. Compared to the other lattice determinations that use the three-point function method, our value is slightly smaller (in magnitude) than the quenched result, $\mu_{\Omega^-} = -1.697 \pm 0.065 $ $\mu_N$, of Boinepalli et.al~\cite{PhysRevD.80.054505} and agrees with the Alexandrou et.al's extrapolated value, $\mu_{\Omega^-} = -1.875 \pm 0.399 $ $\mu_N$~\cite{Alexandrou:2010jv} within one sigma error. In Ref.\cite{PhysRevD.79.051502} magnetic moment of $\Omega$ has determined to be $\mu_{\Omega^-} = -1.93 \pm 0.08 $, by a background field method on $m_\pi=366$ MeV lattices. 

Magnetic moments of $\Omega_c$ and $\Omega_c^\ast$ are very close to each other suggesting that the spin flip of the charm quark has a small effect, as one would expect from a heavy-quark spin symmetry perspective. Based on a quark-model interpretation one would expect the magnetic moment of $\Omega_c$ ($\Omega_{cc}$) to be similar in magnitude to that of $\Omega_c^\ast$ ($\Omega_{cc}^\ast$)'s. While in the case of $\Omega_c$ our finding is consistent with such an expectation, the magnetic moments of $\Omega_{cc}$ and $\Omega_{cc}^\ast$ differ drastically, the latter having a completely vanishing magnetic moment. The difference between the $\Omega_c$ and $\Omega_c^\ast$ is that the $c$-quark is anti-aligned with the $ss$ component in $\Omega_c$ whereas it is aligned in $\Omega_c^\ast$. Combined with their electric charges, quark sectors add constructively for $\Omega_c$ and destructively for $\Omega_c^\ast$. These two different behaviours occur in such a balanced way that the magnetic moments of the $\Omega_c$ and $\Omega_c^\ast$ end up to be similar. In case of the doubly-charmed $\Omega_{cc}$ and $\Omega_{cc}^\ast$ however, the interplay between the electric charges and the number of quarks breaks the balance and lead to magnetic moments for $\Omega_{cc}$ and $\Omega_{cc}^\ast$ that differ significantly.        
\subsubsection*{Findings for the magnetic moments}
\begin{itemize}
	\item Quark sector contributions amongst the spin-3/2 baryons are similar to each other, consonant with the quark-model expectations. \\
	\item Magnetic moments of the strange and charm quarks in spin-3/2 charmed baryons are larger than spin-1/2 baryons having a similar quark-flavor composition. \\
	\item Magnetic moment of the $\Omega^-$ baryon is found to be, $\mu_{\Omega^-} = -1.533 \pm 0.055$ $\mu_N$. \\
	\item $\Omega_c$, $\Omega_c^\ast$ have similar magnetic moments in magnitude. \\
	\item Magnetic moment of the $\Omega_{cc}^\ast$ vanishes unlike the $\Omega_{cc}$. \\
	\item As compared to the light decuplet sector, strange-quark contributions to the magnetic moments of spin-3/2 charmed baryons are smaller.
\end{itemize}
\subsection{Electric-quadrupole form factors}
The electric-quadrupole form factors of spin-3/2 baryons provide information about the deviation of the baryon shape from spherical symmetry. In the Breit frame, the quadrupole form factor and the electric charge distribution are related as~\cite{PhysRevD.46.3067},
\begin{equation}
	\mathcal{G}_{E2}(0) = M^2_B \int d^3 r \bar{\psi}(r) (3z^2 - r^2) \psi (r),
\end{equation}
where $3z^2 - r$ is the standard operator used for quadrupole moments. A positively charged baryon has a prolate (oblate) charge distribution when quadrupole form factor is positive (negative). 

As in the case of the $E0$ and $M1$ form factors, we estimate the $E2$ form factor by the plateau approach. We compute and extract the $s$- and $c$-quark sector contributions individually. $E2$ form factors in lattice units are shown in Fig.~\ref{Fig:E2_plato} and the numerical values, in units of $e/M_B^2$, are given in Table~\ref{E2_res_table}.
\begin{table}[t]
	\caption{$E2(Q^2)$ results for the $\Omega$, $\Omega_c^\ast$, $\Omega_{cc}^\ast$ and $\Omega_{ccc}$ at $\mathbf{q}^2=0.183$ GeV$^2$. Values are given in units of $[e/m^2]$. Quark sector contributions are for single quark and normalised to unit charge. Last column is calculated by the Eq.~\ref{obs_comb}.}
	\label{E2_res_table}
\begin{tabularx}{0.5\textwidth}{l|CCC}
		\hline\hline
													& $E_2(Q^2)_s$ 	& $E_2(Q^2)_c$ 	& $E_2(Q^2)$ \\
		\hline\hline
													& $[e/m^2]$      	& $[e/m^2]$      	& $[e/m^2]$ \\		
		$\Omega$ 			 		    & -0.337(1.142) 	& ---		      		& 0.337(1.142) \\
		$\Omega_c^\ast$ 			& -0.371(539)   	& -0.577(269)			& -0.137(352)  \\
		$\Omega_{cc}^\ast$ 		& -0.091(277)   	& -0.255(87) 		& -0.310(128)  \\
		$\Omega_{ccc}$ 			 	& ---		        	& -0.136(38)  		&	-0.273(76)    \\
		\hline\hline
		
	\end{tabularx}
\end{table}
Unfortunately, low signal/noise ratio does not allow us to conclude about $\Omega$ and $\Omega_c^\ast$ baryons. In the case of the heavier $\Omega_{cc}^\ast$ and $\Omega_{ccc}$ baryons, however, the statistical precision is conclusive and it is possible to make a prediction about their shapes. $\Omega_{cc}^{\ast+}$ and $\Omega_{ccc}^{++}$ have negative $E2$ moments thus their charge distributions deform to an \emph{oblate} shape.
\subsubsection*{Findings for the quadrupole form factors}
\begin{itemize}
	\item $\Omega_{cc}^{\ast+}$ and $\Omega_{ccc}^{++}$ have oblate charge distribution.
\end{itemize}

\subsection{Notes on systematics}

\begin{enumerate}
	\item It is known that the Clover action has $\mathcal{O}(am_q)$ discretisation errors and therefore the simulations may suffer from uncontrolled systematic errors when this action is employed for heavy quarks such as the charm quark. In Ref.~\cite{Can:2013zpa} we estimated the effect of the discretisation errors arising from valence Clover charm quarks in a doubly-heavy $\Xi_{cc}$ system to be small. Since we employ a similar action and formalism in the current work, we expect the discretisation errors to be small. However it is useful to check the significance of such systematic errors once again with the current data set. We have repeated our simulations with parameter $\kappa_c = 0.1256$ which leads to a decrease in the spin-3/2 charmed baryon masses by approximately $3\%$ MeV from those in Table~\ref{bar_mass}. Nevertheless such a change in the charm-quark hopping parameter affects the $E0$ and $M1$ form factors by less than $1 \%$ and $\sim3\%$, respectively. This change is propagated to the electric charge radii and magnetic moments as a change by less than $3\%$. As an illustrative case, we quote the results for $\Omega_c^\ast$ and $\Omega_{ccc}$ baryons. We find $\langle r_E^2 ,\Omega_{ccc} \rangle_{\kappa_c=0.1246} = 0.170(9)$ fm$^2$ as compared to $\langle r_E^2 ,\Omega_{ccc}  \rangle_{\kappa_c=0.1256} = 0.175(10)$ fm$^2$ and $\mu_{\Omega_c^\ast}^{\kappa_c=0.1246} = -0.696(50) \, \mu_N$ as compared to $\mu_{\Omega_c^\ast}^{\kappa_c = 0.1256} = -0.712(50) \, \mu_N$ for 43 measurements. We note that this exercise gives only an indirect probe of discretization errors. A dedicated analysis with different lattice spacings is needed to quantify to which extent our results are prone to such errors.
	\item The set of gauge configurations we have used in this work has an $m_\pi L = 2.3 $ value below the empirical bound of 4, for which the finite volume effects are considered to be significant. In order to quantify the finite-volume effects, one should repeat the computation on lattices with different spatial extent while keeping the $m_\pi$ constant. Since this approach is currently beyond our computational ability, we resort to changing the $m_\pi$ while keeping the spatial extent constant. Rather than a thorough check, this approach is suitable to inspect the severity of the finite-size effects. We compare our $\Omega_c$ and $\Omega_{cc}$ results obtained on $\kappa^{u,d}=0.13781$ lattices to the ones we have calculated previously on $\kappa^{u,d}=0.13770$ and $0.13754$ lattices~\cite{Can:2013tna} for which $m_\pi L = 4.3$ and $m_\pi L = 6$ respectively. Numerical results are given in Table~\ref{mpiL_table} and shown in Fig.~\ref{Fig:mpiL} for a clear comparison. Agreement within error bars suggests that the finite size effects are not severe for the charmed baryons although the $m_\pi L = 2.3$ is well below the empirical value. 
\begin{figure}[ht]
	\centering
	\includegraphics[width=0.45\textwidth]{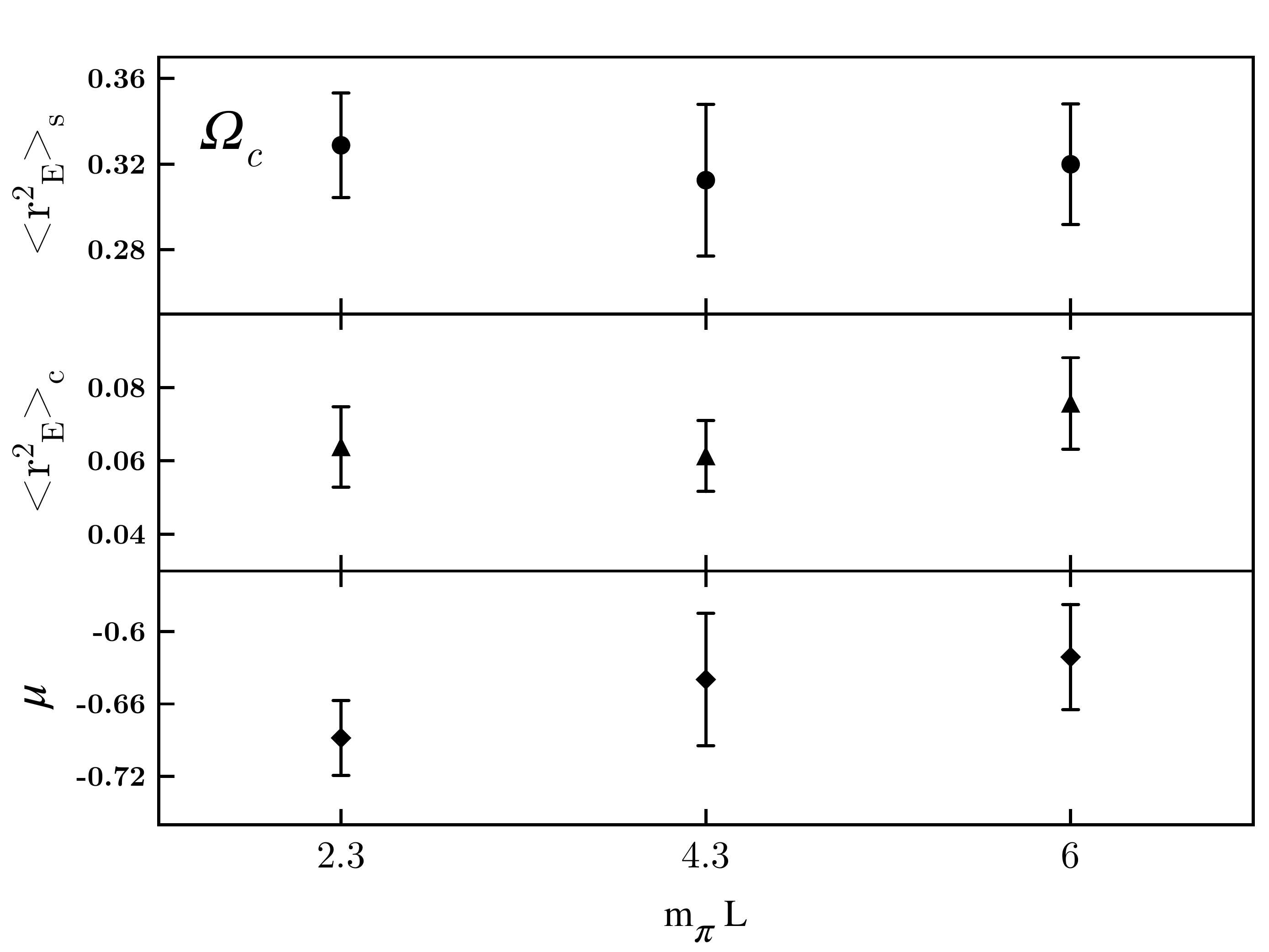}
	\includegraphics[width=0.45\textwidth]{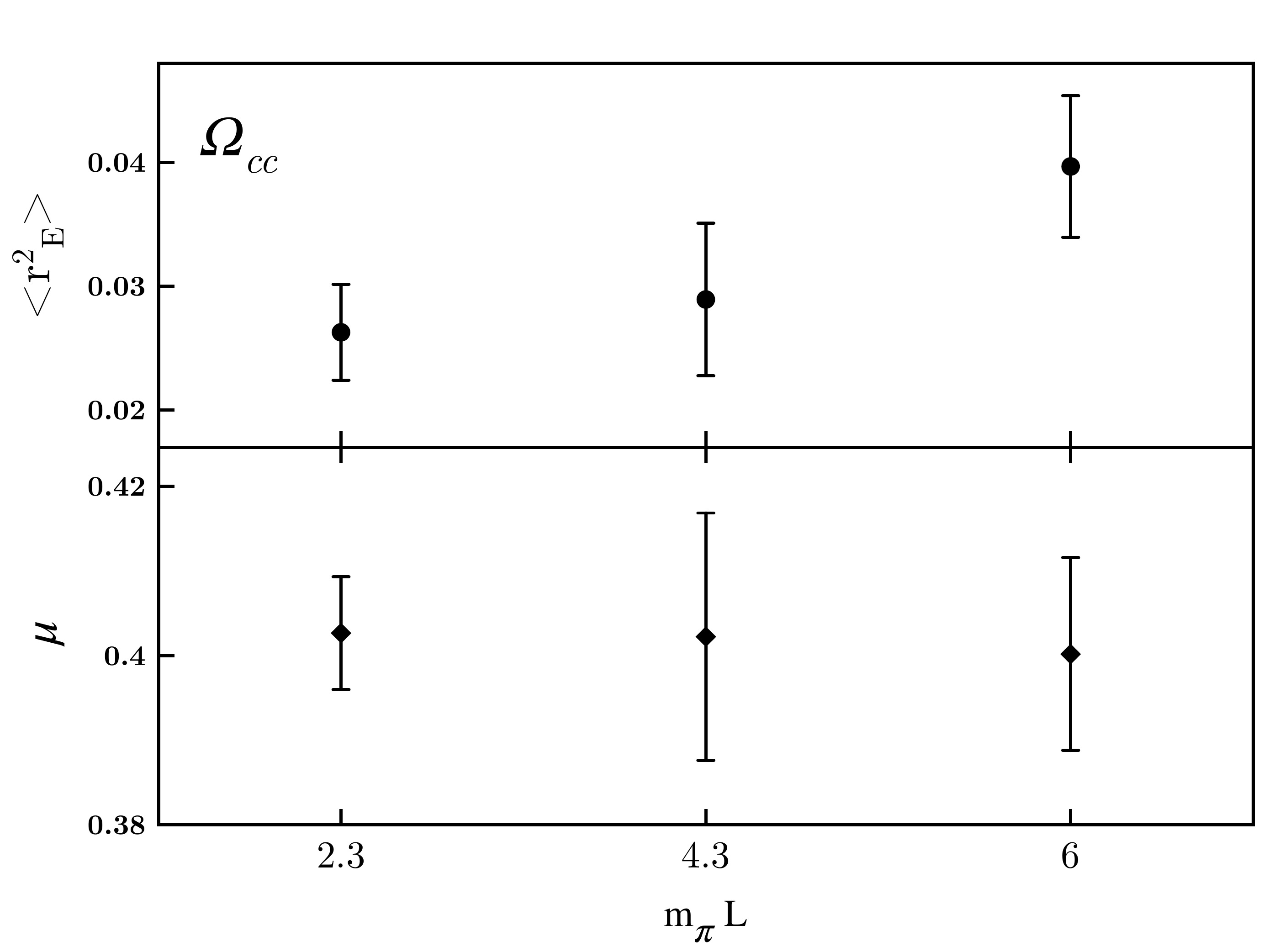}	
	\caption{Electric charge radius and magnetic moment of $\Omega_c$ and $\Omega_{cc}$ baryons on different $m_\pi L$ values. Since the $\Omega_c$ baryon has zero net electric charge we give the strange and charm quark charge radii. }
	\label{Fig:mpiL}
\end{figure}

\begin{table}[ht]
	\caption{Electric charge radius and magnetic moment of $\Omega_c$ and $\Omega_{cc}$ baryons on different $m_\pi L$ values.}
	\label{mpiL_table}
	\begin{tabularx}{0.5\textwidth}{lr|CCC}
		\hline\hline
										& $m_\pi L$ & 2.3 	& 4.3 	& 6 \\
		\hline\hline
										& $\langle r^2_E\rangle_s$ [fm$^2$] & 0.329(24) & 0.313(36) & 0.320(28) \\
		$\Omega_c$ 			& $\langle r^2_E\rangle_c$ [fm$^2$] & 0.064(11) & 0.061(10) & 0.076(13) \\
										& $\mu$ [$\mu_N$] & -0.688(31) & -0.640(55) & -0.621(44) \\
		\hline\hline										
		$\Omega_{cc}$   & $\langle r^2_E\rangle$ [fm$^2$] & 0.026(4) & 0.029(6) & 0.040(6) \\	
										& $\mu$ [$\mu_N$] & 0.403(7) & 0.402(15) & 0.400(11) \\														
		\hline\hline

	\end{tabularx}
\end{table}
	\item In order to extract the charge radii and magnetic moments of the spin-3/2 baryons, we use only two sets of momentum values, namely $[Q^2=0, Q^2=1]$, and employ Eqs.~\eqref{chrgR_dipole} and \eqref{scaling}. A more frequently used, alternative approach is to perform a full form-factor fit on higher Euclidean momentum region. The method we use allows us to estimate the observables of spin-3/2 baryons more precisely since the higher momentum insertions introduce larger statistical errors. Although a fit to a set of $Q^2$ values is desirable to estimate the observables, the values extracted by Eq.~\eqref{chrgR_dipole} or Eq.~\eqref{scaling} agree with those obtained by a form factor fit. As an illustrative example we can compare our values for the spin-1/2 $\Omega_c$ baryon given in Table~\ref{chrgR_dipole_table}. For the spin-1/2 baryons we perform a dipole fit to their form factors as outlined in Ref.~\cite{Can:2013tna}, which are quoted in the middle column. The values given in the rightmost column are extracted using Eq.~\eqref{chrgR_dipole} for comparison. Considering the current precision, we conclude that both approaches agree.
\begin{table}[ht]
	\caption{Electric charge radius of $\Omega_c$ extracted by a dipole fit to the electric form factor and by use of Eq.\eqref{chrgR_dipole}.}
	\label{chrgR_dipole_table}
	\begin{tabularx}{0.5\textwidth}{r|CC}
		\hline\hline
		 																		& Dipole fit 	& Eq.\eqref{chrgR_dipole} \\
		\hline\hline
		$\langle r^2_E\rangle_s$ [fm$^2$] 	& 0.329(24) 	& 0.338(26) \\
		$\langle r^2_E\rangle_c$ [fm$^2$] 	& 0.064(11) 	& 0.067(11) \\
		$\langle r^2_E\rangle$ [fm$^2$] 		& -0.177(18) 	& -0.180(20) \\
		\hline\hline

	\end{tabularx}
\end{table}
	\item Another source of systematic error is due to different fit strategies. One can either perform fits to the data sets of individual quark sectors and then combine the fit results via Eq.~\eqref{obs_comb} to construct the baryon properties or, first, the data sets of the quark sectors can be combined to perform a fit to extract the baryon properties directly. In principle both methods should lead to the same result, however a difference may arise due to statistical fluctuations. Such fluctuations cancel when data sets which are correlated with each other are combined, leading to a better estimated value especially if the value is close to zero like in the case of the $\langle r^2_E\rangle_{\Omega_{cc}}$. When possible we choose to follow this strategy to perform our fits, e.g. spin-1/2 values quoted in Tables~\ref{E_res_table} and \ref{MM_res_table}. One caveat is that, in the case when the electric charge of the baryon is zero, combining data sets naturally leads to a zero electric charge radius. In such cases (e.g. $\langle r^2_E\rangle_{\Omega_c}$) we do not extract the observable from a combined data set. One can check whether two approaches agree with each other by comparing for example $\mu_{\Omega_c} = -0.688(31)$ $\mu_N$ from Table.~\ref{MM_res_table} (fit to combination of quark sectors) with $\mu_{\Omega_c} = -0.714(35)$ $\mu_N$ as obtained via Eq.~\eqref{obs_comb} (fit to individual quark sectors). For the spin-3/2 sector, since we do not have more than two $Q^2$ values, a form factor fit is not possible. Then we combine the fits to the data sets of individual quark sectors by using Eq.~\eqref{obs_comb}.
	\item When computing the matrix elements, one must optimize the temporal separation of the source and sink so as to minimize the excited-state contamination but also to obtain a good signal with minimum amount of statistics. For the electromagnetic form factors it has been shown that a separation of around 1 fm is sufficient~\cite{Can:2013tna, Alexandrou:2011db, Alexandrou:2010hf, Alexandrou:2010jv}. One of the strategies to extract less contaminated values is the so-called \emph{summed operator insertions} (SOI) method~\cite{Maiani:1987by}. In the usual plateau method a generic ratio has the form,
	\begin{align}
		\begin{split}
			R(t_2,t_1;\mathbf{p^\prime},\mathbf{p};\mathbf{\Gamma}) = R_G &+ \mathcal{O}\left(e^{-\Delta t_1}\right) \\
			&+ \mathcal{O} \left(e^{\Delta^\prime(t_2 - t_1)} \right),
		\end{split}
	\end{align} 
where the $R_G$ is the ground state value of the ratio and the first excited state contributions are suppressed proportional to $t_1$ and $(t_2 - t_1)$. $\Delta$ and $\Delta^\prime$ are the energy gap between the ground and first excited state of the source and sink baryons respectively. In the SOI method, one sums the ratio in $t_1$ up to $t_2$ so that it assumes the form,
 	\begin{align}
		\begin{split}
	 		\sum_{t_1=0}^{t_2}R(t_2,t_1;&\mathbf{p^\prime},\mathbf{p};\mathbf{\Gamma}) = R_G.t_2 + c(\Delta, \Delta^\prime) \\
			&+ \mathcal{O}\left(t_2 e^{-\Delta t_2}\right) + \mathcal{O} \left(t_2 e^{\Delta^\prime t_2} \right),
		\end{split}
 	\end{align}
where the $c(\Delta,\Delta^\prime)$ is a constant. First excited-state contributions are now suppressed by $t_2$, which is larger than $t_1$ or $(t_2 - t_1)$. It is possible to calculate the ratio with different source-sink separations and extract the ground state value $R_G$ from the slope of a linear function in $t_2$. Heavy-quark spin symmetry suggests that as the mass of the quarks increase, the energy gap between the ground and excited states of the baryons decrease. In Fig.~\ref{Fig:SOI} we show the $E0$ form factors of $\Omega_{cc}^\ast$ and $\Omega_{ccc}$ baryons, for which we expect the excited state contamination to be the severest. A comparison between the electric charge radius values extracted by plateau and SOI methods are given in Table~\ref{plvssoi_table}. Agreement between the results suggests that the excited-state contamination is under control for $t=1.09$ fm separation.
\begin{figure}[ht]
	\centering
	\includegraphics[width=0.45\textwidth]{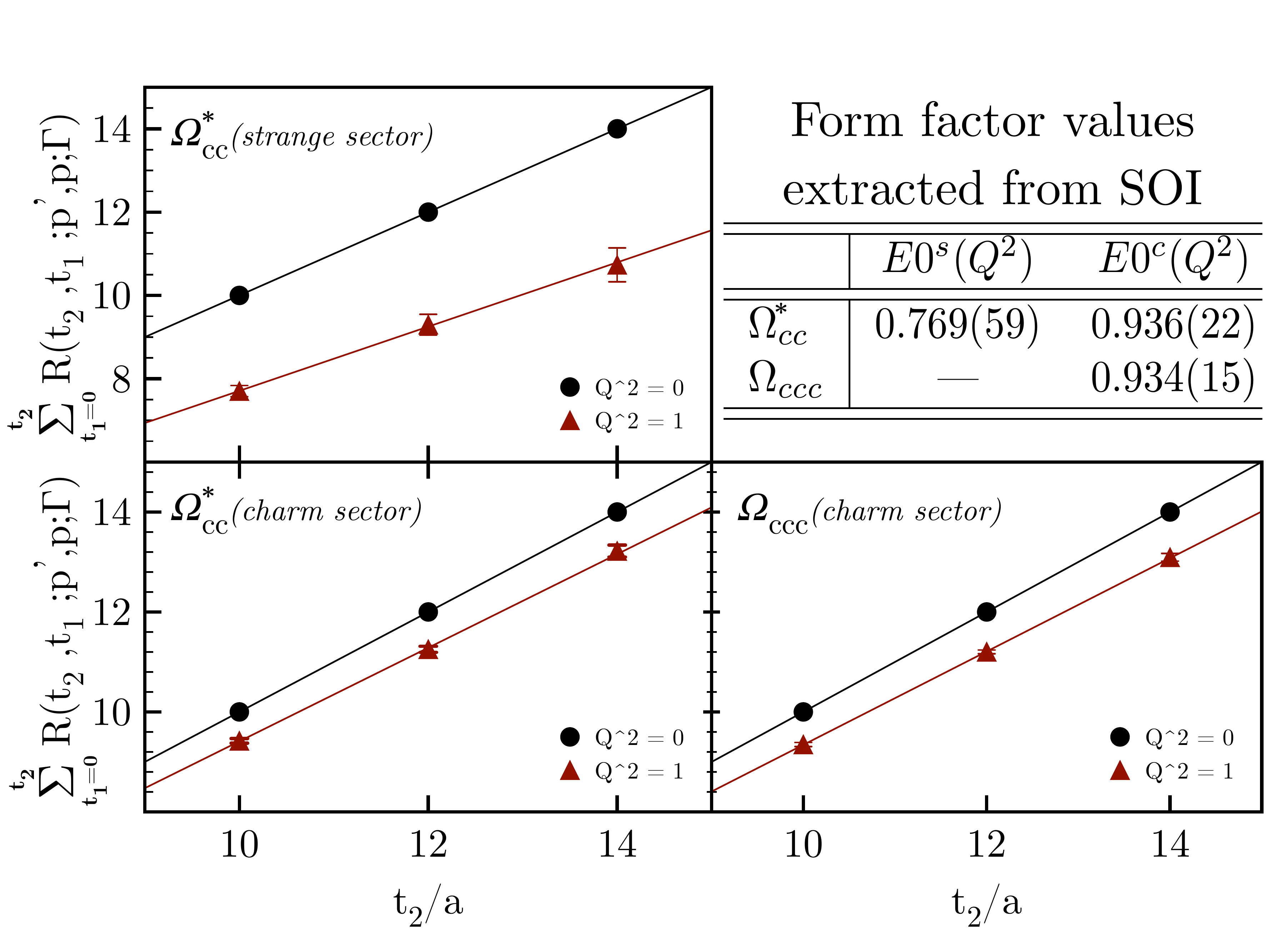}	
	\caption{$E0$ form factors of $\Omega_{cc}^\ast$ and $\Omega_{ccc}$ baryons for two momentum insertions as obtained from summed operator insertions method. Results are for 43 measurements. Values given in the table are for $Q^2=1$. }
	\label{Fig:SOI}
\end{figure}

\begin{table}[ht]
	\caption{Electric charge radius of $\Omega_{cc}^\ast$ and $\Omega_{ccc}$ baryons extracted by the plateau and the SOI method. Results are compared for 43 measurements.}
	\label{plvssoi_table}
	\begin{tabularx}{0.5\textwidth}{r|CC|CC}
		\hline\hline
																			& \multicolumn{2}{c|}{$\Omega_{cc}^\ast$} & \multicolumn{2}{c}{$\Omega_{ccc}$}	\\																			
																			& Plateau 				& SOI 									& Plateau 				& SOI 							\\
		\hline\hline
		$\langle r^2_E\rangle_s$ [fm$^2$] & 0.332(43) 			& 0.359(113) 						& ---				 			& ---				 				\\
		$\langle r^2_E\rangle_c$ [fm$^2$] & 0.079(8) 				& 0.086(31) 						& 0.085(5) 				& 0.089(21) 				\\
		$\langle r^2_E\rangle$ [fm$^2$] 	& -0.005(16) 			& -0.006(53) 						& 0.170(9) 				& 0.179(42) 				\\
		\hline\hline
		
	\end{tabularx}
\end{table}
 \end{enumerate}
 
\section{Summary}
\label{conc}
We have calculated the electromagnetic form factors of the $\Omega$, $\Omega_c^\ast$, $\Omega_{cc}^\ast$ and $\Omega_{ccc}$ baryons at the lowest allowed three-momentum value ($\mathbf{q}^2=0.183$ GeV$^2$) on the lattices we use, and extracted their electric charge radii, magnetic moments and quadrupole moments. Based on the method outlined in our previous work~\cite{Can:2013tna}, we have computed the electromagnetic form factors of the $\Omega_c$ and $\Omega_{cc}$ baryons also and extracted the electric charge radii and magnetic moments. For each observable we have identified the quark sector contributions as well as the baryon properties. 

We find that the electric charge radii of the strange sector is insensitive to the composition of the baryon, whereas the charm sector shows a slight dependence to the number of the charm quarks that compose the baryon. Spin flip has a significant effect on the doubly-represented quark sectors so that the charge radii of the doubly-represented strange and charm quarks increase when the spin of the singly-represented quark is flipped.

In case of the magnetic moments, quark-sector behaviours change drastically between the spin-1/2 and spin-3/2 charmed baryons such that the quark-sector contributions to the  magnetic moments of spin-3/2 baryons get enhanced. Magnetic moments of $\Omega_c$ and $\Omega_c^\ast$ baryons are found to be similar indicating a negligible spin-flip effect by the singly-represented charm quark. $\Omega_{cc}^\ast$ baryon has a vanishing magnetic moment unlike the spin-1/2 $\Omega_{cc}$ baryon. Strange-quark contributions to the magnetic moments decrease within the charmed baryons compared to the decuplet sector.     

Finally, we have been able to achieve a statistically significant data for the quadrupole moments of the $\Omega_{cc}^\ast$ and $\Omega_{ccc}$ baryons to conclude that their electric charge distributions deform to oblate shape.   
   
\acknowledgments
Part of numerical calculations in this work were performed on National Center for High Performance Computing of Turkey (Istanbul Technical University) under project number 10462009. The unquenched gauge configurations employed in our analysis were generated by PACS-CS collaboration~\cite{Aoki:2008sm}. We used a modified version of Chroma software system~\cite{Edwards:2004sx} along with QUDA~\cite{Babich:2011np,Clark:2009wm}. This work is supported in part by The Scientiﬁc and Technological Research Council of Turkey (TUBITAK) under project number 114F261 and in part by KAKENHI under Contract Nos. 25247036 and 24250294. This work is also supported by the Research Abroad and Invitational Program for the Promotion of International Joint Research, Category (C) and the International Physics Leadership Program at Tokyo Tech. G. Erkol would like to thank M. Oka and Tokyo Institute of Technology, where part of this work was done, for the hospitality they provided.


\begin{thebibliography}{34}%
\makeatletter
\providecommand \@ifxundefined [1]{%
 \@ifx{#1\undefined}
}%
\providecommand \@ifnum [1]{%
 \ifnum #1\expandafter \@firstoftwo
 \else \expandafter \@secondoftwo
 \fi
}%
\providecommand \@ifx [1]{%
 \ifx #1\expandafter \@firstoftwo
 \else \expandafter \@secondoftwo
 \fi
}%
\providecommand \natexlab [1]{#1}%
\providecommand \enquote  [1]{``#1''}%
\providecommand \bibnamefont  [1]{#1}%
\providecommand \bibfnamefont [1]{#1}%
\providecommand \citenamefont [1]{#1}%
\providecommand \href@noop [0]{\@secondoftwo}%
\providecommand \href [0]{\begingroup \@sanitize@url \@href}%
\providecommand \@href[1]{\@@startlink{#1}\@@href}%
\providecommand \@@href[1]{\endgroup#1\@@endlink}%
\providecommand \@sanitize@url [0]{\catcode `\\12\catcode `\$12\catcode
  `\&12\catcode `\#12\catcode `\^12\catcode `\_12\catcode `\%12\relax}%
\providecommand \@@startlink[1]{}%
\providecommand \@@endlink[0]{}%
\providecommand \url  [0]{\begingroup\@sanitize@url \@url }%
\providecommand \@url [1]{\endgroup\@href {#1}{\urlprefix }}%
\providecommand \urlprefix  [0]{URL }%
\providecommand \Eprint [0]{\href }%
\providecommand \doibase [0]{http://dx.doi.org/}%
\providecommand \selectlanguage [0]{\@gobble}%
\providecommand \bibinfo  [0]{\@secondoftwo}%
\providecommand \bibfield  [0]{\@secondoftwo}%
\providecommand \translation [1]{[#1]}%
\providecommand \BibitemOpen [0]{}%
\providecommand \bibitemStop [0]{}%
\providecommand \bibitemNoStop [0]{.\EOS\space}%
\providecommand \EOS [0]{\spacefactor3000\relax}%
\providecommand \BibitemShut  [1]{\csname bibitem#1\endcsname}%
\let\auto@bib@innerbib\@empty
\bibitem [{\citenamefont {Constantinou}(2015)}]{Constantinou:2014tga}%
  \BibitemOpen
  \bibfield  {author} {\bibinfo {author} {\bibfnamefont {M.}~\bibnamefont
  {Constantinou}},\ }\href@noop {} {\bibfield  {journal} {\bibinfo  {journal}
  {PoS}\ }\textbf {\bibinfo {volume} {LATTICE2014}},\ \bibinfo {pages} {001}
  (\bibinfo {year} {2015})},\ \Eprint {http://arxiv.org/abs/1411.0078}
  {arXiv:1411.0078 [hep-lat]} \BibitemShut {NoStop}%
\bibitem [{\citenamefont {Boinepalli}\ \emph {et~al.}(2006)\citenamefont
  {Boinepalli}, \citenamefont {Leinweber}, \citenamefont {Williams},
  \citenamefont {Zanotti},\ and\ \citenamefont {Zhang}}]{PhysRevD.74.093005}%
  \BibitemOpen
  \bibfield  {author} {\bibinfo {author} {\bibfnamefont {S.}~\bibnamefont
  {Boinepalli}}, \bibinfo {author} {\bibfnamefont {D.~B.}\ \bibnamefont
  {Leinweber}}, \bibinfo {author} {\bibfnamefont {A.~G.}\ \bibnamefont
  {Williams}}, \bibinfo {author} {\bibfnamefont {J.~M.}\ \bibnamefont
  {Zanotti}}, \ and\ \bibinfo {author} {\bibfnamefont {J.~B.}\ \bibnamefont
  {Zhang}},\ }\href {\doibase 10.1103/PhysRevD.74.093005} {\bibfield  {journal}
  {\bibinfo  {journal} {Phys. Rev. D}\ }\textbf {\bibinfo {volume} {74}},\
  \bibinfo {pages} {093005} (\bibinfo {year} {2006})}\BibitemShut {NoStop}%
\bibitem [{\citenamefont {Shanahan}\ \emph
  {et~al.}(2014{\natexlab{a}})\citenamefont {Shanahan}, \citenamefont
  {Horsley}, \citenamefont {Nakamura}, \citenamefont {Pleiter}, \citenamefont
  {Rakow}, \citenamefont {Schierholz}, \citenamefont {Stuben}, \citenamefont
  {Thomas}, \citenamefont {Young},\ and\ \citenamefont
  {Zanotti}}]{Shanahan:2014uka}%
  \BibitemOpen
  \bibfield  {author} {\bibinfo {author} {\bibfnamefont {P.~E.}\ \bibnamefont
  {Shanahan}}, \bibinfo {author} {\bibfnamefont {R.}~\bibnamefont {Horsley}},
  \bibinfo {author} {\bibfnamefont {Y.}~\bibnamefont {Nakamura}}, \bibinfo
  {author} {\bibfnamefont {D.}~\bibnamefont {Pleiter}}, \bibinfo {author}
  {\bibfnamefont {P.~E.~L.}\ \bibnamefont {Rakow}}, \bibinfo {author}
  {\bibfnamefont {G.}~\bibnamefont {Schierholz}}, \bibinfo {author}
  {\bibfnamefont {H.}~\bibnamefont {Stuben}}, \bibinfo {author} {\bibfnamefont
  {A.~W.}\ \bibnamefont {Thomas}}, \bibinfo {author} {\bibfnamefont {R.~D.}\
  \bibnamefont {Young}}, \ and\ \bibinfo {author} {\bibfnamefont {J.~M.}\
  \bibnamefont {Zanotti}} (\bibinfo {collaboration} {CSSM Collaboration, QCDSF/UKQCD Collaboration}),\
  }\href {\doibase 10.1103/PhysRevD.89.074511} {\bibfield  {journal} {\bibinfo
  {journal} {Phys.Rev.}\ }\textbf {\bibinfo {volume} {D89}},\ \bibinfo {pages}
  {074511} (\bibinfo {year} {2014}{\natexlab{a}})},\ \Eprint
  {http://arxiv.org/abs/1401.5862} {arXiv:1401.5862 [hep-lat]} \BibitemShut
  {NoStop}%
\bibitem [{\citenamefont {Shanahan}\ \emph
  {et~al.}(2014{\natexlab{b}})\citenamefont {Shanahan}, \citenamefont {Thomas},
  \citenamefont {Young}, \citenamefont {Zanotti}, \citenamefont {Horsley} \emph
  {et~al.}}]{Shanahan:2014cga}%
  \BibitemOpen
  \bibfield  {author} {\bibinfo {author} {\bibfnamefont {P.~E.}\ \bibnamefont
  {Shanahan}}, \bibinfo {author} {\bibfnamefont {A.~W.}\ \bibnamefont
  {Thomas}}, \bibinfo {author} {\bibfnamefont {R.~D.}\ \bibnamefont {Young}},
  \bibinfo {author} {\bibfnamefont {J.~M.}\ \bibnamefont {Zanotti}}, \bibinfo
  {author} {\bibfnamefont {R.}~\bibnamefont {Horsley}},  \emph {et~al.},\
  }\href {\doibase 10.1103/PhysRevD.90.034502} {\bibfield  {journal} {\bibinfo
  {journal} {Phys.Rev.}\ }\textbf {\bibinfo {volume} {D90}},\ \bibinfo {pages}
  {034502} (\bibinfo {year} {2014}{\natexlab{b}})},\ \Eprint
  {http://arxiv.org/abs/1403.1965} {arXiv:1403.1965 [hep-lat]} \BibitemShut
  {NoStop}%
\bibitem [{\citenamefont {Boinepalli}\ \emph {et~al.}(2009)\citenamefont
  {Boinepalli}, \citenamefont {Leinweber}, \citenamefont {Moran}, \citenamefont
  {Williams}, \citenamefont {Zanotti},\ and\ \citenamefont
  {Zhang}}]{PhysRevD.80.054505}%
  \BibitemOpen
  \bibfield  {author} {\bibinfo {author} {\bibfnamefont {S.}~\bibnamefont
  {Boinepalli}}, \bibinfo {author} {\bibfnamefont {D.~B.}\ \bibnamefont
  {Leinweber}}, \bibinfo {author} {\bibfnamefont {P.~J.}\ \bibnamefont
  {Moran}}, \bibinfo {author} {\bibfnamefont {A.~G.}\ \bibnamefont {Williams}},
  \bibinfo {author} {\bibfnamefont {J.~M.}\ \bibnamefont {Zanotti}}, \ and\
  \bibinfo {author} {\bibfnamefont {J.~B.}\ \bibnamefont {Zhang}},\ }\href
  {\doibase 10.1103/PhysRevD.80.054505} {\bibfield  {journal} {\bibinfo
  {journal} {Phys. Rev. D}\ }\textbf {\bibinfo {volume} {80}},\ \bibinfo
  {pages} {054505} (\bibinfo {year} {2009})}\BibitemShut {NoStop}%
\bibitem [{\citenamefont {Alexandrou}\ \emph {et~al.}(2009)\citenamefont
  {Alexandrou}, \citenamefont {Korzec}, \citenamefont {Koutsou}, \citenamefont
  {Leontiou}, \citenamefont {Lorce} \emph {et~al.}}]{Alexandrou:2008bn}%
  \BibitemOpen
  \bibfield  {author} {\bibinfo {author} {\bibfnamefont {C.}~\bibnamefont
  {Alexandrou}}, \bibinfo {author} {\bibfnamefont {T.}~\bibnamefont {Korzec}},
  \bibinfo {author} {\bibfnamefont {G.}~\bibnamefont {Koutsou}}, \bibinfo
  {author} {\bibfnamefont {T.}~\bibnamefont {Leontiou}}, \bibinfo {author}
  {\bibfnamefont {C.}~\bibnamefont {Lorce}},  \emph {et~al.},\ }\href {\doibase
  10.1103/PhysRevD.79.014507} {\bibfield  {journal} {\bibinfo  {journal}
  {Phys.Rev.}\ }\textbf {\bibinfo {volume} {D79}},\ \bibinfo {pages} {014507}
  (\bibinfo {year} {2009})},\ \Eprint {http://arxiv.org/abs/0810.3976}
  {arXiv:0810.3976 [hep-lat]} \BibitemShut {NoStop}%
\bibitem [{\citenamefont {Alexandrou}\ \emph {et~al.}(2010)\citenamefont
  {Alexandrou}, \citenamefont {Korzec}, \citenamefont {Koutsou}, \citenamefont
  {Negele},\ and\ \citenamefont {Proestos}}]{Alexandrou:2010jv}%
  \BibitemOpen
  \bibfield  {author} {\bibinfo {author} {\bibfnamefont {C.}~\bibnamefont
  {Alexandrou}}, \bibinfo {author} {\bibfnamefont {T.}~\bibnamefont {Korzec}},
  \bibinfo {author} {\bibfnamefont {G.}~\bibnamefont {Koutsou}}, \bibinfo
  {author} {\bibfnamefont {J.~W.}\ \bibnamefont {Negele}}, \ and\ \bibinfo
  {author} {\bibfnamefont {Y.}~\bibnamefont {Proestos}},\ }\href {\doibase
  10.1103/PhysRevD.82.034504} {\bibfield  {journal} {\bibinfo  {journal}
  {Phys.Rev.}\ }\textbf {\bibinfo {volume} {D82}},\ \bibinfo {pages} {034504}
  (\bibinfo {year} {2010})},\ \Eprint {http://arxiv.org/abs/1006.0558}
  {arXiv:1006.0558 [hep-lat]} \BibitemShut {NoStop}%
\bibitem [{\citenamefont {Can}\ \emph {et~al.}(2013{\natexlab{a}})\citenamefont
  {Can}, \citenamefont {Erkol}, \citenamefont {Oka}, \citenamefont {Ozpineci},\
  and\ \citenamefont {Takahashi}}]{Can:2012tx}%
  \BibitemOpen
  \bibfield  {author} {\bibinfo {author} {\bibfnamefont {K.}~\bibnamefont
  {Can}}, \bibinfo {author} {\bibfnamefont {G.}~\bibnamefont {Erkol}}, \bibinfo
  {author} {\bibfnamefont {M.}~\bibnamefont {Oka}}, \bibinfo {author}
  {\bibfnamefont {A.}~\bibnamefont {Ozpineci}}, \ and\ \bibinfo {author}
  {\bibfnamefont {T.}~\bibnamefont {Takahashi}},\ }\href {\doibase
  10.1016/j.physletb.2012.12.050} {\bibfield  {journal} {\bibinfo  {journal}
  {Phys.Lett.}\ }\textbf {\bibinfo {volume} {B719}},\ \bibinfo {pages} {103}
  (\bibinfo {year} {2013}{\natexlab{a}})},\ \Eprint
  {http://arxiv.org/abs/1210.0869} {arXiv:1210.0869 [hep-lat]} \BibitemShut
  {NoStop}%
\bibitem [{\citenamefont {Can}\ \emph {et~al.}(2013{\natexlab{b}})\citenamefont
  {Can}, \citenamefont {Erkol}, \citenamefont {Isildak}, \citenamefont {Oka},\
  and\ \citenamefont {Takahashi}}]{Can:2013zpa}%
  \BibitemOpen
  \bibfield  {author} {\bibinfo {author} {\bibfnamefont {K.}~\bibnamefont
  {Can}}, \bibinfo {author} {\bibfnamefont {G.}~\bibnamefont {Erkol}}, \bibinfo
  {author} {\bibfnamefont {B.}~\bibnamefont {Isildak}}, \bibinfo {author}
  {\bibfnamefont {M.}~\bibnamefont {Oka}}, \ and\ \bibinfo {author}
  {\bibfnamefont {T.}~\bibnamefont {Takahashi}},\ }\href {\doibase
  10.1016/j.physletb.2013.09.024} {\bibfield  {journal} {\bibinfo  {journal}
  {Phys.Lett.}\ }\textbf {\bibinfo {volume} {B726}},\ \bibinfo {pages} {703}
  (\bibinfo {year} {2013}{\natexlab{b}})},\ \Eprint
  {http://arxiv.org/abs/1306.0731} {arXiv:1306.0731 [hep-lat]} \BibitemShut
  {NoStop}%
\bibitem [{\citenamefont {Can}\ \emph {et~al.}(2014)\citenamefont {Can},
  \citenamefont {Erkol}, \citenamefont {Isildak}, \citenamefont {Oka},\ and\
  \citenamefont {Takahashi}}]{Can:2013tna}%
  \BibitemOpen
  \bibfield  {author} {\bibinfo {author} {\bibfnamefont {K.}~\bibnamefont
  {Can}}, \bibinfo {author} {\bibfnamefont {G.}~\bibnamefont {Erkol}}, \bibinfo
  {author} {\bibfnamefont {B.}~\bibnamefont {Isildak}}, \bibinfo {author}
  {\bibfnamefont {M.}~\bibnamefont {Oka}}, \ and\ \bibinfo {author}
  {\bibfnamefont {T.}~\bibnamefont {Takahashi}},\ }\href {\doibase
  10.1007/JHEP05(2014)125} {\bibfield  {journal} {\bibinfo  {journal} {JHEP}\
  }\textbf {\bibinfo {volume} {1405}},\ \bibinfo {pages} {125} (\bibinfo {year}
  {2014})},\ \Eprint {http://arxiv.org/abs/1310.5915} {arXiv:1310.5915
  [hep-lat]} \BibitemShut {NoStop}%
\bibitem [{\citenamefont {Bahtiyar}\ \emph {et~al.}(2015)\citenamefont
  {Bahtiyar}, \citenamefont {Can}, \citenamefont {Erkol},\ and\ \citenamefont
  {Oka}}]{Bahtiyar2015281}%
  \BibitemOpen
  \bibfield  {author} {\bibinfo {author} {\bibfnamefont {H.}~\bibnamefont
  {Bahtiyar}}, \bibinfo {author} {\bibfnamefont {K.}~\bibnamefont {Can}},
  \bibinfo {author} {\bibfnamefont {G.}~\bibnamefont {Erkol}}, \ and\ \bibinfo
  {author} {\bibfnamefont {M.}~\bibnamefont {Oka}},\ }\href {\doibase
  http://dx.doi.org/10.1016/j.physletb.2015.06.006} {\bibfield  {journal}
  {\bibinfo  {journal} {Physics Letters B}\ }\textbf {\bibinfo {volume}
  {747}},\ \bibinfo {pages} {281 } (\bibinfo {year} {2015})}\BibitemShut
  {NoStop}%
\bibitem [{\citenamefont {Nozawa}\ and\ \citenamefont
  {Leinweber}(1990)}]{Nozawa:1990gt}%
  \BibitemOpen
  \bibfield  {author} {\bibinfo {author} {\bibfnamefont {S.}~\bibnamefont
  {Nozawa}}\ and\ \bibinfo {author} {\bibfnamefont {D.~B.}\ \bibnamefont
  {Leinweber}},\ }\href {\doibase 10.1103/PhysRevD.42.3567} {\bibfield
  {journal} {\bibinfo  {journal} {Phys.Rev.}\ }\textbf {\bibinfo {volume}
  {D42}},\ \bibinfo {pages} {3567} (\bibinfo {year} {1990})}\BibitemShut
  {NoStop}%
\bibitem [{\citenamefont {Alexandrou}\ \emph {et~al.}(2014)\citenamefont
  {Alexandrou}, \citenamefont {Drach}, \citenamefont {Jansen}, \citenamefont
  {Kallidonis},\ and\ \citenamefont {Koutsou}}]{Alexandrou:2014sha}%
  \BibitemOpen
  \bibfield  {author} {\bibinfo {author} {\bibfnamefont {C.}~\bibnamefont
  {Alexandrou}}, \bibinfo {author} {\bibfnamefont {V.}~\bibnamefont {Drach}},
  \bibinfo {author} {\bibfnamefont {K.}~\bibnamefont {Jansen}}, \bibinfo
  {author} {\bibfnamefont {C.}~\bibnamefont {Kallidonis}}, \ and\ \bibinfo
  {author} {\bibfnamefont {G.}~\bibnamefont {Koutsou}},\ }\href {\doibase
  10.1103/PhysRevD.90.074501} {\bibfield  {journal} {\bibinfo  {journal}
  {Phys.Rev.}\ }\textbf {\bibinfo {volume} {D90}},\ \bibinfo {pages} {074501}
  (\bibinfo {year} {2014})},\ \Eprint {http://arxiv.org/abs/1406.4310}
  {arXiv:1406.4310 [hep-lat]} \BibitemShut {NoStop}%
\bibitem [{\citenamefont {Aoki}\ \emph {et~al.}(2009)\citenamefont {Aoki},
  \citenamefont {Ishikawa}, \citenamefont {Ishizuka}, \citenamefont {Izubuchi},
  \citenamefont {Kadoh}, \citenamefont {Kanaya}, \citenamefont {Kuramashi},
  \citenamefont {Namekawa}, \citenamefont {Okawa}, \citenamefont {Taniguchi},
  \citenamefont {Ukawa}, \citenamefont {Ukita},\ and\ \citenamefont
  {Yoshie}}]{Aoki:2008sm}%
  \BibitemOpen
  \bibfield  {author} {\bibinfo {author} {\bibfnamefont {S.}~\bibnamefont
  {Aoki}}, \bibinfo {author} {\bibfnamefont {K.-I.}\ \bibnamefont {Ishikawa}},
  \bibinfo {author} {\bibfnamefont {N.}~\bibnamefont {Ishizuka}}, \bibinfo
  {author} {\bibfnamefont {T.}~\bibnamefont {Izubuchi}}, \bibinfo {author}
  {\bibfnamefont {D.}~\bibnamefont {Kadoh}}, \bibinfo {author} {\bibfnamefont
  {K.}~\bibnamefont {Kanaya}}, \bibinfo {author} {\bibfnamefont
  {Y.}~\bibnamefont {Kuramashi}}, \bibinfo {author} {\bibfnamefont
  {Y.}~\bibnamefont {Namekawa}}, \bibinfo {author} {\bibfnamefont
  {M.}~\bibnamefont {Okawa}}, \bibinfo {author} {\bibfnamefont
  {Y.}~\bibnamefont {Taniguchi}}, \bibinfo {author} {\bibfnamefont
  {A.}~\bibnamefont {Ukawa}}, \bibinfo {author} {\bibfnamefont
  {N.}~\bibnamefont {Ukita}}, \ and\ \bibinfo {author} {\bibfnamefont
  {T.}~\bibnamefont {Yoshie}} (\bibinfo {collaboration} {PACS-CS Collaboration}),\ }\href
  {\doibase 10.1103/PhysRevD.79.034503} {\bibfield  {journal} {\bibinfo
  {journal} {Phys. Rev.}\ }\textbf {\bibinfo {volume} {D79}},\ \bibinfo {pages}
  {034503} (\bibinfo {year} {2009})},\ \Eprint {http://arxiv.org/abs/0807.1661}
  {arXiv:0807.1661 [hep-lat]} \BibitemShut {NoStop}%
\bibitem [{\citenamefont {El-Khadra}\ \emph {et~al.}(1997)\citenamefont
  {El-Khadra}, \citenamefont {Kronfeld},\ and\ \citenamefont
  {Mackenzie}}]{ElKhadra:1996mp}%
  \BibitemOpen
  \bibfield  {author} {\bibinfo {author} {\bibfnamefont {A.~X.}\ \bibnamefont
  {El-Khadra}}, \bibinfo {author} {\bibfnamefont {A.~S.}\ \bibnamefont
  {Kronfeld}}, \ and\ \bibinfo {author} {\bibfnamefont {P.~B.}\ \bibnamefont
  {Mackenzie}},\ }\href {\doibase 10.1103/PhysRevD.55.3933} {\bibfield
  {journal} {\bibinfo  {journal} {Phys. Rev.}\ }\textbf {\bibinfo {volume}
  {D55}},\ \bibinfo {pages} {3933} (\bibinfo {year} {1997})},\ \Eprint
  {http://arxiv.org/abs/hep-lat/9604004} {arXiv:hep-lat/9604004} \BibitemShut
  {NoStop}%
\bibitem [{\citenamefont {Burch}\ \emph {et~al.}(2010)\citenamefont {Burch},
  \citenamefont {DeTar}, \citenamefont {Di~Pierro}, \citenamefont {El-Khadra},
  \citenamefont {Freeland} \emph {et~al.}}]{Burch:2009az}%
  \BibitemOpen
  \bibfield  {author} {\bibinfo {author} {\bibfnamefont {T.}~\bibnamefont
  {Burch}}, \bibinfo {author} {\bibfnamefont {C.}~\bibnamefont {DeTar}},
  \bibinfo {author} {\bibfnamefont {M.}~\bibnamefont {Di~Pierro}}, \bibinfo
  {author} {\bibfnamefont {A.}~\bibnamefont {El-Khadra}}, \bibinfo {author}
  {\bibfnamefont {E.}~\bibnamefont {Freeland}},  \emph {et~al.},\ }\href
  {\doibase 10.1103/PhysRevD.81.034508} {\bibfield  {journal} {\bibinfo
  {journal} {Phys.Rev.}\ }\textbf {\bibinfo {volume} {D81}},\ \bibinfo {pages}
  {034508} (\bibinfo {year} {2010})},\ \Eprint {http://arxiv.org/abs/0912.2701}
  {arXiv:0912.2701 [hep-lat]} \BibitemShut {NoStop}%
\bibitem [{\citenamefont {Bernard}\ \emph {et~al.}(2011)\citenamefont
  {Bernard}, \citenamefont {DeTar}, \citenamefont {DiPierro}, \citenamefont
  {El-Khadra}, \citenamefont {Evans}, \citenamefont {Freeland}, \citenamefont
  {Gamiz}, \citenamefont {Gottlieb}, \citenamefont {Heller}, \citenamefont
  {Hetrick}, \citenamefont {Kronfeld}, \citenamefont {Laiho}, \citenamefont
  {Levkova}, \citenamefont {Mackenzie}, \citenamefont {Simone}, \citenamefont
  {Sugar}, \citenamefont {Toussaint},\ and\ \citenamefont
  {VandeWater}}]{Bernard:2010fr}%
  \BibitemOpen
  \bibfield  {author} {\bibinfo {author} {\bibfnamefont {C.}~\bibnamefont
  {Bernard}}, \bibinfo {author} {\bibfnamefont {C.}~\bibnamefont {DeTar}},
  \bibinfo {author} {\bibfnamefont {M.}~\bibnamefont {DiPierro}}, \bibinfo
  {author} {\bibfnamefont {A.~X.}\ \bibnamefont {El-Khadra}}, \bibinfo {author}
  {\bibfnamefont {R.~T.}\ \bibnamefont {Evans}}, \bibinfo {author}
  {\bibfnamefont {E.~D.}\ \bibnamefont {Freeland}}, \bibinfo {author}
  {\bibfnamefont {E.}~\bibnamefont {Gamiz}}, \bibinfo {author} {\bibfnamefont
  {S.}~\bibnamefont {Gottlieb}}, \bibinfo {author} {\bibfnamefont {U.~M.}\
  \bibnamefont {Heller}}, \bibinfo {author} {\bibfnamefont {J.~E.}\
  \bibnamefont {Hetrick}}, \bibinfo {author} {\bibfnamefont {A.~S.}\
  \bibnamefont {Kronfeld}}, \bibinfo {author} {\bibfnamefont {J.}~\bibnamefont
  {Laiho}}, \bibinfo {author} {\bibfnamefont {L.}~\bibnamefont {Levkova}},
  \bibinfo {author} {\bibfnamefont {P.~B.}\ \bibnamefont {Mackenzie}}, \bibinfo
  {author} {\bibfnamefont {J.~N.}\ \bibnamefont {Simone}}, \bibinfo {author}
  {\bibfnamefont {R.}~\bibnamefont {Sugar}}, \bibinfo {author} {\bibfnamefont
  {D.}~\bibnamefont {Toussaint}}, \ and\ \bibinfo {author} {\bibfnamefont
  {R.~S.}\ \bibnamefont {VandeWater}} (\bibinfo {collaboration} {Fermilab
  Lattice Collaboration, MILC Collaboration}),\ }\href {\doibase
  10.1103/PhysRevD.83.034503} {\bibfield  {journal} {\bibinfo  {journal}
  {Phys.Rev.}\ }\textbf {\bibinfo {volume} {D83}},\ \bibinfo {pages} {034503}
  (\bibinfo {year} {2011})},\ \Eprint {http://arxiv.org/abs/1003.1937}
  {arXiv:1003.1937 [hep-lat]} \BibitemShut {NoStop}%
\bibitem [{\citenamefont {Mohler}\ and\ \citenamefont
  {Woloshyn}(2011)}]{Mohler:2011ke}%
  \BibitemOpen
  \bibfield  {author} {\bibinfo {author} {\bibfnamefont {D.}~\bibnamefont
  {Mohler}}\ and\ \bibinfo {author} {\bibfnamefont {R.~M.}\ \bibnamefont
  {Woloshyn}},\ }\href {\doibase 10.1103/PhysRevD.84.054505} {\bibfield
  {journal} {\bibinfo  {journal} {Phys.Rev.}\ }\textbf {\bibinfo {volume}
  {D84}},\ \bibinfo {pages} {054505} (\bibinfo {year} {2011})},\ \Eprint
  {http://arxiv.org/abs/1103.5506} {arXiv:1103.5506 [hep-lat]} \BibitemShut
  {NoStop}%
\bibitem [{\citenamefont {Mohler}\ \emph
  {et~al.}(2013{\natexlab{a}})\citenamefont {Mohler}, \citenamefont
  {Prelovsek},\ and\ \citenamefont {Woloshyn}}]{Mohler:2012na}%
  \BibitemOpen
  \bibfield  {author} {\bibinfo {author} {\bibfnamefont {D.}~\bibnamefont
  {Mohler}}, \bibinfo {author} {\bibfnamefont {S.}~\bibnamefont {Prelovsek}}, \
  and\ \bibinfo {author} {\bibfnamefont {R.~M.}\ \bibnamefont {Woloshyn}},\
  }\href {\doibase 10.1103/PhysRevD.87.034501} {\bibfield  {journal} {\bibinfo
  {journal} {Phys.Rev.}\ }\textbf {\bibinfo {volume} {D87}},\ \bibinfo {pages}
  {034501} (\bibinfo {year} {2013}{\natexlab{a}})},\ \Eprint
  {http://arxiv.org/abs/1208.4059} {arXiv:1208.4059 [hep-lat]} \BibitemShut
  {NoStop}%
\bibitem [{\citenamefont {Mohler}\ \emph
  {et~al.}(2013{\natexlab{b}})\citenamefont {Mohler}, \citenamefont {Lang},
  \citenamefont {Leskovec}, \citenamefont {Prelovsek},\ and\ \citenamefont
  {Woloshyn}}]{Mohler:2013rwa}%
  \BibitemOpen
  \bibfield  {author} {\bibinfo {author} {\bibfnamefont {D.}~\bibnamefont
  {Mohler}}, \bibinfo {author} {\bibfnamefont {C.~B.}\ \bibnamefont {Lang}},
  \bibinfo {author} {\bibfnamefont {L.}~\bibnamefont {Leskovec}}, \bibinfo
  {author} {\bibfnamefont {S.}~\bibnamefont {Prelovsek}}, \ and\ \bibinfo
  {author} {\bibfnamefont {R.~M.}\ \bibnamefont {Woloshyn}},\ }\href@noop {} {\
   (\bibinfo {year} {2013}{\natexlab{b}})},\ \Eprint
  {http://arxiv.org/abs/1308.3175} {arXiv:1308.3175 [hep-lat]} \BibitemShut
  {NoStop}%
\bibitem [{\citenamefont {Aoki}\ \emph {et~al.}(1996)\citenamefont {Aoki},
  \citenamefont {Fukugita}, \citenamefont {Hashimoto}, \citenamefont {Iwasaki},
  \citenamefont {Kanaya}, \citenamefont {Kuramashi}, \citenamefont {Mino},
  \citenamefont {Okawa}, \citenamefont {Ukawa},\ and\ \citenamefont
  {Yoshie}}]{Aoki:1995bb}%
  \BibitemOpen
  \bibfield  {author} {\bibinfo {author} {\bibfnamefont {S.}~\bibnamefont
  {Aoki}}, \bibinfo {author} {\bibfnamefont {M.}~\bibnamefont {Fukugita}},
  \bibinfo {author} {\bibfnamefont {S.}~\bibnamefont {Hashimoto}}, \bibinfo
  {author} {\bibfnamefont {Y.}~\bibnamefont {Iwasaki}}, \bibinfo {author}
  {\bibfnamefont {K.}~\bibnamefont {Kanaya}}, \bibinfo {author} {\bibfnamefont
  {Y.}~\bibnamefont {Kuramashi}}, \bibinfo {author} {\bibfnamefont
  {H.}~\bibnamefont {Mino}}, \bibinfo {author} {\bibfnamefont {M.}~\bibnamefont
  {Okawa}}, \bibinfo {author} {\bibfnamefont {A.}~\bibnamefont {Ukawa}}, \ and\
  \bibinfo {author} {\bibfnamefont {T.}~\bibnamefont {Yoshie}} (\bibinfo
  {collaboration} {JLQCD Collaboration}),\ }\href {\doibase
  10.1016/0920-5632(96)00072-2} {\bibfield  {journal} {\bibinfo  {journal}
  {Nucl. Phys. Proc. Suppl.}\ }\textbf {\bibinfo {volume} {47}},\ \bibinfo
  {pages} {354} (\bibinfo {year} {1996})},\ \Eprint
  {http://arxiv.org/abs/hep-lat/9510013} {arXiv:hep-lat/9510013 [hep-lat]}
  \BibitemShut {NoStop}%
\bibitem [{\citenamefont {Namekawa}\ \emph {et~al.}(2013)\citenamefont
  {Namekawa}, \citenamefont {Aoki}, \citenamefont {Ishikawa}, \citenamefont
  {Ishizuka}, \citenamefont {Kanaya}, \citenamefont {Kuramashi}, \citenamefont
  {Okawa}, \citenamefont {Taniguchi}, \citenamefont {Ukawa}, \citenamefont
  {Ukita},\ and\ \citenamefont {Yoshie}}]{Namekawa:2013vu}%
  \BibitemOpen
  \bibfield  {author} {\bibinfo {author} {\bibfnamefont {Y.}~\bibnamefont
  {Namekawa}}, \bibinfo {author} {\bibfnamefont {S.}~\bibnamefont {Aoki}},
  \bibinfo {author} {\bibfnamefont {K.~I.}\ \bibnamefont {Ishikawa}}, \bibinfo
  {author} {\bibfnamefont {N.}~\bibnamefont {Ishizuka}}, \bibinfo {author}
  {\bibfnamefont {K.}~\bibnamefont {Kanaya}}, \bibinfo {author} {\bibfnamefont
  {Y.}~\bibnamefont {Kuramashi}}, \bibinfo {author} {\bibfnamefont
  {M.}~\bibnamefont {Okawa}}, \bibinfo {author} {\bibfnamefont
  {Y.}~\bibnamefont {Taniguchi}}, \bibinfo {author} {\bibfnamefont
  {A.}~\bibnamefont {Ukawa}}, \bibinfo {author} {\bibfnamefont
  {N.}~\bibnamefont {Ukita}}, \ and\ \bibinfo {author} {\bibfnamefont
  {T.}~\bibnamefont {Yoshie}} (\bibinfo {collaboration} {PACS-CS
  Collaboration}),\ }\href {\doibase 10.1103/PhysRevD.87.094512} {\bibfield
  {journal} {\bibinfo  {journal} {Phys.Rev.}\ }\textbf {\bibinfo {volume}
  {D87}},\ \bibinfo {pages} {094512} (\bibinfo {year} {2013})},\ \Eprint
  {http://arxiv.org/abs/1301.4743} {arXiv:1301.4743 [hep-lat]} \BibitemShut
  {NoStop}%
\bibitem [{\citenamefont {Briceno}\ \emph {et~al.}(2012)\citenamefont
  {Briceno}, \citenamefont {Lin},\ and\ \citenamefont
  {Bolton}}]{Briceno:2012wt}%
  \BibitemOpen
  \bibfield  {author} {\bibinfo {author} {\bibfnamefont {R.~A.}\ \bibnamefont
  {Briceno}}, \bibinfo {author} {\bibfnamefont {H.-W.}\ \bibnamefont {Lin}}, \
  and\ \bibinfo {author} {\bibfnamefont {D.~R.}\ \bibnamefont {Bolton}},\
  }\href {\doibase 10.1103/PhysRevD.86.094504} {\bibfield  {journal} {\bibinfo
  {journal} {Phys.Rev.}\ }\textbf {\bibinfo {volume} {D86}},\ \bibinfo {pages}
  {094504} (\bibinfo {year} {2012})},\ \Eprint {http://arxiv.org/abs/1207.3536}
  {arXiv:1207.3536 [hep-lat]} \BibitemShut {NoStop}%
\bibitem [{\citenamefont {Brown}\ \emph {et~al.}(2014)\citenamefont {Brown},
  \citenamefont {Detmold}, \citenamefont {Meinel},\ and\ \citenamefont
  {Orginos}}]{Brown:2014ena}%
  \BibitemOpen
  \bibfield  {author} {\bibinfo {author} {\bibfnamefont {Z.~S.}\ \bibnamefont
  {Brown}}, \bibinfo {author} {\bibfnamefont {W.}~\bibnamefont {Detmold}},
  \bibinfo {author} {\bibfnamefont {S.}~\bibnamefont {Meinel}}, \ and\ \bibinfo
  {author} {\bibfnamefont {K.}~\bibnamefont {Orginos}},\ }\href {\doibase
  10.1103/PhysRevD.90.094507} {\bibfield  {journal} {\bibinfo  {journal} {Phys.
  Rev.}\ }\textbf {\bibinfo {volume} {D90}},\ \bibinfo {pages} {094507}
  (\bibinfo {year} {2014})},\ \Eprint {http://arxiv.org/abs/1409.0497}
  {arXiv:1409.0497 [hep-lat]} \BibitemShut {NoStop}%
\bibitem [{\citenamefont {Olive}\ \emph {et~al.}(2014)\citenamefont {Olive}
  \emph {et~al.}}]{Agashe:2014kda}%
  \BibitemOpen
  \bibfield  {author} {\bibinfo {author} {\bibfnamefont {K.}~\bibnamefont
  {Olive}} \emph {et~al.} (\bibinfo {collaboration} {Particle Data Group}),\
  }\href {\doibase 10.1088/1674-1137/38/9/090001} {\bibfield  {journal}
  {\bibinfo  {journal} {Chin.Phys.}\ }\textbf {\bibinfo {volume} {C38}},\
  \bibinfo {pages} {090001} (\bibinfo {year} {2014})}\BibitemShut {NoStop}%
\bibitem [{\citenamefont {Leinweber}\ \emph {et~al.}(1993)\citenamefont
  {Leinweber}, \citenamefont {Draper},\ and\ \citenamefont
  {Woloshyn}}]{Leinweber:1992pv}%
  \BibitemOpen
  \bibfield  {author} {\bibinfo {author} {\bibfnamefont {D.~B.}\ \bibnamefont
  {Leinweber}}, \bibinfo {author} {\bibfnamefont {T.}~\bibnamefont {Draper}}, \
  and\ \bibinfo {author} {\bibfnamefont {R.~M.}\ \bibnamefont {Woloshyn}},\
  }\href {\doibase 10.1103/PhysRevD.48.2230} {\bibfield  {journal} {\bibinfo
  {journal} {Phys.Rev.}\ }\textbf {\bibinfo {volume} {D48}},\ \bibinfo {pages}
  {2230} (\bibinfo {year} {1993})},\ \Eprint
  {http://arxiv.org/abs/hep-lat/9212016} {arXiv:hep-lat/9212016 [hep-lat]}
  \BibitemShut {NoStop}%
\bibitem [{\citenamefont {Aubin}\ \emph {et~al.}(2009)\citenamefont {Aubin},
  \citenamefont {Orginos}, \citenamefont {Pascalutsa},\ and\ \citenamefont
  {Vanderhaeghen}}]{PhysRevD.79.051502}%
  \BibitemOpen
  \bibfield  {author} {\bibinfo {author} {\bibfnamefont {C.}~\bibnamefont
  {Aubin}}, \bibinfo {author} {\bibfnamefont {K.}~\bibnamefont {Orginos}},
  \bibinfo {author} {\bibfnamefont {V.}~\bibnamefont {Pascalutsa}}, \ and\
  \bibinfo {author} {\bibfnamefont {M.}~\bibnamefont {Vanderhaeghen}},\ }\href
  {\doibase 10.1103/PhysRevD.79.051502} {\bibfield  {journal} {\bibinfo
  {journal} {Phys. Rev. D}\ }\textbf {\bibinfo {volume} {79}},\ \bibinfo
  {pages} {051502} (\bibinfo {year} {2009})}\BibitemShut {NoStop}%
\bibitem [{\citenamefont {Leinweber}\ \emph {et~al.}(1992)\citenamefont
  {Leinweber}, \citenamefont {Draper},\ and\ \citenamefont
  {Woloshyn}}]{PhysRevD.46.3067}%
  \BibitemOpen
  \bibfield  {author} {\bibinfo {author} {\bibfnamefont {D.~B.}\ \bibnamefont
  {Leinweber}}, \bibinfo {author} {\bibfnamefont {T.}~\bibnamefont {Draper}}, \
  and\ \bibinfo {author} {\bibfnamefont {R.~M.}\ \bibnamefont {Woloshyn}},\
  }\href {\doibase 10.1103/PhysRevD.46.3067} {\bibfield  {journal} {\bibinfo
  {journal} {Phys. Rev. D}\ }\textbf {\bibinfo {volume} {46}},\ \bibinfo
  {pages} {3067} (\bibinfo {year} {1992})}\BibitemShut {NoStop}%
\bibitem [{\citenamefont {Alexandrou}\ \emph
  {et~al.}(2011{\natexlab{a}})\citenamefont {Alexandrou}, \citenamefont
  {Brinet}, \citenamefont {Carbonell}, \citenamefont {Constantinou},
  \citenamefont {Harraud} \emph {et~al.}}]{Alexandrou:2011db}%
  \BibitemOpen
  \bibfield  {author} {\bibinfo {author} {\bibfnamefont {C.}~\bibnamefont
  {Alexandrou}}, \bibinfo {author} {\bibfnamefont {M.}~\bibnamefont {Brinet}},
  \bibinfo {author} {\bibfnamefont {J.}~\bibnamefont {Carbonell}}, \bibinfo
  {author} {\bibfnamefont {M.}~\bibnamefont {Constantinou}}, \bibinfo {author}
  {\bibfnamefont {P.}~\bibnamefont {Harraud}},  \emph {et~al.},\ }\href
  {\doibase 10.1103/PhysRevD.83.094502} {\bibfield  {journal} {\bibinfo
  {journal} {Phys.Rev.}\ }\textbf {\bibinfo {volume} {D83}},\ \bibinfo {pages}
  {094502} (\bibinfo {year} {2011}{\natexlab{a}})},\ \Eprint
  {http://arxiv.org/abs/1102.2208} {arXiv:1102.2208 [hep-lat]} \BibitemShut
  {NoStop}%
\bibitem [{\citenamefont {Alexandrou}\ \emph
  {et~al.}(2011{\natexlab{b}})\citenamefont {Alexandrou}, \citenamefont
  {Brinet}, \citenamefont {Carbonell}, \citenamefont {Constantinou},
  \citenamefont {Harraud}, \citenamefont {Guichon}, \citenamefont {Jansen},
  \citenamefont {Korzec},\ and\ \citenamefont {Papinutto}}]{Alexandrou:2010hf}%
  \BibitemOpen
  \bibfield  {author} {\bibinfo {author} {\bibfnamefont {C.}~\bibnamefont
  {Alexandrou}}, \bibinfo {author} {\bibfnamefont {M.}~\bibnamefont {Brinet}},
  \bibinfo {author} {\bibfnamefont {J.}~\bibnamefont {Carbonell}}, \bibinfo
  {author} {\bibfnamefont {M.}~\bibnamefont {Constantinou}}, \bibinfo {author}
  {\bibfnamefont {P.~A.}\ \bibnamefont {Harraud}}, \bibinfo {author}
  {\bibfnamefont {P.}~\bibnamefont {Guichon}}, \bibinfo {author} {\bibfnamefont
  {K.}~\bibnamefont {Jansen}}, \bibinfo {author} {\bibfnamefont
  {T.}~\bibnamefont {Korzec}}, \ and\ \bibinfo {author} {\bibfnamefont
  {M.}~\bibnamefont {Papinutto}} (\bibinfo {collaboration} {ETM Collaboration}),\ }\href
  {\doibase 10.1103/PhysRevD.83.045010} {\bibfield  {journal} {\bibinfo
  {journal} {Phys. Rev.}\ }\textbf {\bibinfo {volume} {D83}},\ \bibinfo {pages}
  {045010} (\bibinfo {year} {2011}{\natexlab{b}})},\ \Eprint
  {http://arxiv.org/abs/1012.0857} {arXiv:1012.0857 [hep-lat]} \BibitemShut
  {NoStop}%
\bibitem [{\citenamefont {Maiani}\ \emph {et~al.}(1987)\citenamefont {Maiani},
  \citenamefont {Martinelli}, \citenamefont {Paciello},\ and\ \citenamefont
  {Taglienti}}]{Maiani:1987by}%
  \BibitemOpen
  \bibfield  {author} {\bibinfo {author} {\bibfnamefont {L.}~\bibnamefont
  {Maiani}}, \bibinfo {author} {\bibfnamefont {G.}~\bibnamefont {Martinelli}},
  \bibinfo {author} {\bibfnamefont {M.~L.}\ \bibnamefont {Paciello}}, \ and\
  \bibinfo {author} {\bibfnamefont {B.}~\bibnamefont {Taglienti}},\ }\href
  {\doibase 10.1016/0550-3213(87)90078-2} {\bibfield  {journal} {\bibinfo
  {journal} {Nucl. Phys.}\ }\textbf {\bibinfo {volume} {B293}},\ \bibinfo
  {pages} {420} (\bibinfo {year} {1987})}\BibitemShut {NoStop}%
\bibitem [{\citenamefont {Edwards}\ and\ \citenamefont
  {Joo}(2005)}]{Edwards:2004sx}%
  \BibitemOpen
  \bibfield  {author} {\bibinfo {author} {\bibfnamefont {R.~G.}\ \bibnamefont
  {Edwards}}\ and\ \bibinfo {author} {\bibfnamefont {B.}~\bibnamefont {Joo}}
  (\bibinfo {collaboration} {SciDAC Collaboration, LHPC Collaboration, UKQCD
  Collaboration}),\ }\href {\doibase 10.1016/j.nuclphysbps.2004.11.254}
  {\bibfield  {journal} {\bibinfo  {journal} {Nucl.Phys.Proc.Suppl.}\ }\textbf
  {\bibinfo {volume} {140}},\ \bibinfo {pages} {832} (\bibinfo {year}
  {2005})},\ \Eprint {http://arxiv.org/abs/hep-lat/0409003}
  {arXiv:hep-lat/0409003 [hep-lat]} \BibitemShut {NoStop}%
\bibitem [{\citenamefont {Babich}\ \emph {et~al.}(2011)\citenamefont {Babich},
  \citenamefont {Clark}, \citenamefont {Joo}, \citenamefont {Shi},
  \citenamefont {Brower} \emph {et~al.}}]{Babich:2011np}%
  \BibitemOpen
  \bibfield  {author} {\bibinfo {author} {\bibfnamefont {R.}~\bibnamefont
  {Babich}}, \bibinfo {author} {\bibfnamefont {M.}~\bibnamefont {Clark}},
  \bibinfo {author} {\bibfnamefont {B.}~\bibnamefont {Joo}}, \bibinfo {author}
  {\bibfnamefont {G.}~\bibnamefont {Shi}}, \bibinfo {author} {\bibfnamefont
  {R.}~\bibnamefont {Brower}},  \emph {et~al.},\ }\href@noop {} {\  (\bibinfo
  {year} {2011})},\ \Eprint {http://arxiv.org/abs/1109.2935} {arXiv:1109.2935
  [hep-lat]} \BibitemShut {NoStop}%
\bibitem [{\citenamefont {Clark}\ \emph {et~al.}(2010)\citenamefont {Clark},
  \citenamefont {Babich}, \citenamefont {Barros}, \citenamefont {Brower},\ and\
  \citenamefont {Rebbi}}]{Clark:2009wm}%
  \BibitemOpen
  \bibfield  {author} {\bibinfo {author} {\bibfnamefont {M.}~\bibnamefont
  {Clark}}, \bibinfo {author} {\bibfnamefont {R.}~\bibnamefont {Babich}},
  \bibinfo {author} {\bibfnamefont {K.}~\bibnamefont {Barros}}, \bibinfo
  {author} {\bibfnamefont {R.}~\bibnamefont {Brower}}, \ and\ \bibinfo {author}
  {\bibfnamefont {C.}~\bibnamefont {Rebbi}},\ }\href {\doibase
  10.1016/j.cpc.2010.05.002} {\bibfield  {journal} {\bibinfo  {journal}
  {Comput.Phys.Commun.}\ }\textbf {\bibinfo {volume} {181}},\ \bibinfo {pages}
  {1517} (\bibinfo {year} {2010})},\ \Eprint {http://arxiv.org/abs/0911.3191}
  {arXiv:0911.3191 [hep-lat]} \BibitemShut {NoStop}%
\end{thebibliography}

%

\end{document}